\documentstyle[eqsecnum,floats,aps,prc,epsfig,twocolumn]{revtex}
\begin{document}

\wideabs{

\title{Crystal Structures of Polymerized Fullerides AC$_{60}$, A=K, Rb, Cs and Alkali-mediated Interactions}
\author{B. Verberck$^1$, K.H. Michel$^1$, and A.V. Nikolaev$^{1,2}$}
\address{
$^1$Department of Physics, University of Antwerp, UIA, 2610 Antwerpen, Belgium\\
$^2$Institute of Physical Chemistry of RAS, Leninskii prospect 31, 117915, Moscow, Russia 
} 
\date{\today} 
\maketitle
%
%--------
%ABSTRACT
%--------
\begin{abstract}
Starting from a model of rigid interacting C$_{60}$ polymer chains on an orthorhombic lattice, we study the
mutual orientation of the chains and the stability of the crystalline structures $Pmnn$ and $I2/m$.  We take
into account i) van der Waals interactions and electric quadrupole interactions between C$_{60}$ monomers on
different chains as well as ii) interactions of the monomers with the surrounding alkali atoms.  The direct
interactions i) always lead to an antiferrorotational structure $Pmnn$ with alternate orientation of the
C$_{60}$ chains in planes $(001)$.  The interactions ii) with the alkalis consist of two parts:
translation-rotation (TR) coupling where the orientations of the chains interact with displacements of the
alkalis, and quadrupolar electronic polarizability (ep) coupling, where the electric quadrupoles on the
C$_{60}$ monomers interact with induced quadrupoles due to excited electronic $d$ states of the alkalis.
Both interactions ii) lead to an effective orientation-orientation interaction between the C$_{60}$ chains
and always favor the ferrorotational structure $I2/m$ where C$_{60}$ chains
have a same orientation.  The structures $Pmnn$ for KC$_{60}$ and $I2/m$ for Rb- and CsC$_{60}$ are the
result of a competition between the direct interaction i) and the alkali-mediated interactions ii).  In Rb-
and CsC$_{60}$ the latter are found to be dominant, the preponderant role being played by the quadrupolar
electronic polarizability of the alkali ions.
\end{abstract}

}

\narrowtext 
 
%
%------------
%INTRODUCTION
%------------
\begin{section}{Introduction}
Alkali metal doped C$_{60}$ (A$_{x}$C$_{60}$), A=K, Rb, Cs, forms stable crystalline phases
(fullerides) with a broad range of physical and chemical properties comprising
superconductors and polymer phases.  For a review, see Refs. \cite{Dre,Kuz,For}.
In particular the $x=1$ compounds \cite{Win} exhibit plastic crystalline phases with cubic
rock-salt structure (space group $Fm\overline{3}m$) at high temperature ($T\geq350$ K) and polymer phases
\cite{Pek,Cha,Ste,Ren} at lower $T$.  In the latter case the C$_{60}$ molecules are linked
through
a [2+2] cycloaddition \cite{Ste}, a mechanism originally proposed for photoinduced polymerization
\cite{Rao} in pristine C$_{60}$.  From X-ray powder diffraction \cite{Ste} it was concluded that
the crystal structure of both KC$_{60}$ and RbC$_{60}$ was orthorhombic (space group $Pmnn$).  Polymerization
occurs along the orthorhombic $\vec{a}$ axis (the former cubic [110] direction).  The orientation of the
polymer
chain, taken as a rigid unit, is characterized by the angle $\psi$ of the planes of cycloaddition with the
orthorhombic $\vec{c}$ axis.  In the $Pmnn$ structure (Fig.\ 1a) the chains have alternating orientations
$+\psi$ and $-\psi$, $\left|\psi\right|\approx45\;^{\rm o}$.
Notwithstanding this apparent structural similarity the electronic and magnetic properties of KC$_{60}$
on one hand and RbC$_{60}$ and CsC$_{60}$ on the other hand were found to be very different \cite{Kuz}.
ESR and optical conductivity data \cite{Cha,Bom} show that RbC$_{60}$ and CsC$_{60}$ exhibit a
transition from a quasi-one-dimensional metal to an insulating magnetic state near 50 K, while KC$_{60}$
stays metallic and nonmagnetic at low $T$.  NMR spectra also did show marked differences between K- and Rb-,
CsC$_{60}$ polymers \cite{All}.  The contradiction between similar crystalline structures and different
electromagnetic
properties was resolved by single crystal X-ray diffraction and diffuse scattering studies \cite{Lau}.
While the space group $Pmnn$ is confirmed for KC$_{60}$, it is found that RbC$_{60}$ is body-centered
monoclinic with space group $I2/m$.  In the latter structure, the polymer chains have the same orientation
$\psi$ (Fig.\ 1b) in successive (001) planes.  High-resolution synchrotron powder
diffraction \cite{Rou} results have demonstrated that CsC$_{60}$ has the same structure as RbC$_{60}$.  Most
recently a detailed structure of the polymer phase of K- and RbC$_{60}$ has been performed by high-resolution
neutron powder diffraction \cite{Huq}.  The distinct crystalline structures are confirmed and in addition
a determination of the positions of the C nuclei demonstrates a distortion of the C$_{60}$ monomers.  The
discovery of a metal-insulator phase transition by ESR spectroscopy in
KC$_{60}$ around $T=50$ K and the concomitant appearance of a superstructure revealed by X-ray diffraction
\cite{Cou} demonstrate a subtle interplay of structure, dimensionality and electronic properties.  In fact
the nature of electrical conductivity in the AC$_{60}$ compounds is still under debate.  Previous electronic
band structure calculations \cite{Erw,Tan} suggest a three-dimensional dispersion of the electronic bands.
However numerous experiments \cite{For} are interpreted in terms of a quasi-one-dimensional conductor.

%-----------------------------------------------------------------------
%     figure 1
%-----------------------------------------------------------------------
\begin{figure} 
\resizebox{0.46\textwidth}{!}
{ 
 \includegraphics{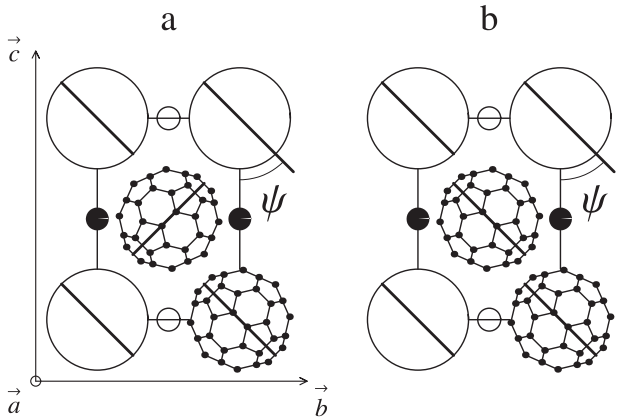} 
} 
\vspace{3mm}
%     figure caption
\caption{
Crystal structures projected onto the crystallographic
 $(\vec{b},\vec{c})$ plane:
(a) $Pmnn$, (b) $I2/m$. The thick bars represent 
the projection of the cycloaddition planes.
Polymerization occurs along $\vec{a}$. 
The alkalis located in $(\vec{b},\vec{c})$ planes and at
$\pm a/2$ are denoted by full $(+)$ and empty $(-)$ circles.
} 
\label{fig1} 
\end{figure} 

Concerning our understanding of the crystalline structure, a condensation scheme  has been
proposed for the phase transition from the orientationally disordered structure $Fm\overline{3}m$
to the orientationally ordered polymer phase $Pmnn$ on basis of a phenomenological
Landau theory \cite{Aks}.  However, a microscopic description of this transition (which involves the
cycloaddition process) is very complex.  Already in the $Fm\overline{3}m$ phase, the charge transfer of
one electron from the alkali atoms to the C$_{60}$ molecules giving rise to a A$^{+}$C$_{60}^{-}$ crystal
changes the electronic structure of the molecule.  This change affects the orientation dependent
interactions, in particular the crystal field, and favors polymerization \cite{Nik}.  The latter is a
quantum-chemical process and within our view it is the driving mechanism for the phase transition.
There remains the question why polymerized KC$_{60}$ and Rb-, CsC$_{60}$ have different crystal structures
and different electronic properties and why pressure polymerized C$_{60}$ has the same crystalline structure
as KC$_{60}$.
In the following we will essentially address the problem of the various crystalline structures.
It turns out that the differences in structure can be related to the average radii of excited electronic
$d$-states of the alkali cations.
A preliminary account has been given as a letter \cite{Mic}.  There it was found that the direct
electric quadrupole interactions between charged polymer chains in the alkali fullerides (the existence of
electric quadrupoles on the chains is a
consequence of the charge transfer between an alkali atom and a C$_{60}$ monomer \cite{Nik2}) always favors
the structure
$Pmnn$ with alternating chain orientations.  While this result agrees with experiment for the case of
KC$_{60}$, it does not explain the monoclinic structure $I2/m$ of RbC$_{60}$ and CsC$_{60}$.  Hence it
becomes necessary to take into account the role of the alkalis.  In particular the specific quadrupolar
polarizability of the alkali metal ions leads to an indirect interchain coupling which favors $I2/m$ with
equal chain orientation.  The competition between direct and indirect interactions then explains the
structural differences between KC$_{60}$ and RbC$_{60}$, CsC$_{60}$.  In the present paper
we will present several important extensions of the previous work.

Since polymerization leads to chains of $D_{2h}$ symmetry, one has to investigate the role of van der Waals
interactions (repulsive Born-Mayer and attractive London dispersion forces) for a configuration that is very
different from the situation in orientationally
disordered C$_{60}$.  Potential energy calculations based on van der Waals forces between C$_{60}$ polymer
chains show that the energy minimum is highly sensitive to the relative chain orientation \cite{Lau2}.
We will show that the dominant term of the multipole expansion of the van der Waals interaction between
different chains, corresponding to quadrupole-quadrupole interactions, always favors the orthorhombic $Pmnn$
structure too.  Our study of the van der Waals interchain interactions
is not only relevant for the structure of the AC$_{60}$ compounds, where the direct interchain interactions
are in competition with the alkali-mediated chain-chain interactions, but also for the understanding of the
structure of polymerized C$_{60}$.  There a ``low pressure" orthorhombic phase was found to have the structure
$Pmnn$ \cite{Dav2,Mor}.

As a further extension we will study the coupling between rotational motion of the polymer chains and the
lattice translations of the alkalis (TR coupling).  Thereby we will obtain the shear mode that characterizes
the monoclinic unit cell of RbC$_{60}$ \cite{Huq} and CsC$_{60}$ \cite{Rou}.  However our calculations show
that the indirect chain-chain interaction mediated by the displacements of the alkalis is relatively weak and
is not able to account for the structural difference between KC$_{60}$ and Rb-, CsC$_{60}$.  The quadrupolar
polarizability mediated interaction is the decisive mechanism.

The content of the paper is as follows.  In Sect.\ \ref{rigid} we formulate the rigid chain model of 
C$_{60}$ polymer chains in the orthorhombic lattice (space group $Immm$).  The sole degree of freedom of each
chain is the rotation angle $\psi$ about the long axis.  Since the interchain potential is a function of the
angles $\psi$, we perform a multipole expansion into symmetry adapted rotator functions (SARFs), taking into
account the symmetry of the polymer chain ($D_{2h}$) and of the site.  The direct interchain potential is
studied for both Coulomb and van der Waals forces between monomers on different chains (Sects.\ 
\ref{rigid} and \ref{Section3}).  These forces lead to Coulomb- and van der Waals type
quadrupole-quadrupole interactions between chains.  Studying the direct interaction potentials in Fourier
space, we find that they always favor the structure $Pmnn$.  Next (Sect.\ \ref{Sectionalkalimediated})
we study the
interactions of the C$_{60}$ chains with the surrounding alkalis.  We consider two types of interactions.
Firstly the coupling of the rotations of the polymers with the alkali displacements (TR-coupling) via van der
Waals and Coulomb forces, and secondly the coupling of rotations of the polymers with induced quadrupoles on
the alkali cations (quadrupolar polarizability coupling).  Both interactions lead to an effective rotation
coupling between polymer chains, which always favors a ferrorotational structure $I2/m$.  In the following,
we study the competition between direct and indirect interactions and discuss the stability of structures
(Sect.\ \ref{Section5}).  Finally (\ref{Sectionconclusions}) conclusions are drawn.  We discuss the unique
properties of
AC$_{60}$ in comparison with ionic molecular crystals with small ions.  The paper has three appendices,
where we discuss: A) the orthorhombic lattice structure as a consequence of polymerization, B) details about
the interchain potentials, C) the microscopic origin of the quadrupolar polarizability of the cations.
\end{section}

%-----------------
%RIGID CHAIN MODEL
%-----------------
\begin{section}{Rigid Chain Model} \label{rigid}
We start from the assumption that the polymerization by stereospecific cycloaddition of C$_{60}$ molecules
in AC$_{60}$ has occurred along the original cubic $[110]$ direction.  Polymerization acts as a negative
uniaxial stress along $[110]$.  By using concepts of the theory of elasticity \cite{Lan}, we find that the
cubic crystal is deformed into an orthorhombic one (point group $D_{2h}$).  Taking the cubic $[110]$,
$[1\overline{1}0]$, and $[001]$ as new $\xi$, $\eta$, $\zeta$ axes (orthorhombic $\vec{a}$, $\vec{b}$,
$\vec{c}$) respectively, we find the deformations
\begin{mathletters}
\begin{eqnarray}
 & & \epsilon_{\xi\xi}=-\frac{K}{4dc_{44}}\left[c_{11}\left(c_{11}+c_{12}+2c_{44}\right)-2c_{12}^2\right], \label{elast1a}\\
 & & \epsilon_{\eta\eta}=\frac{K}{4dc_{44}}\left[c_{11}\left(c_{11}+c_{12}-2c_{44}\right)-2c_{12}^2\right], \label{elast1b}\\
 & & \epsilon_{\zeta\zeta}=Kc_{12}/d, \label{elast1c}
\end{eqnarray}
\end{mathletters}
where $d=c_{11}\left(c_{11}+c_{12}\right)-2c_{12}^2$, $K>0$.  Details of the derivation are given in
Appendix~\ref{AppendixA}.  From the relations between the cubic elastic
constants we see that $\epsilon_{\xi\xi}<0$, $\epsilon_{\eta\eta}>0$ and $\epsilon_{\zeta\zeta}>0$, which
corresponds to a contraction along the $\xi$ direction (which we identify with the orthorhombic $\vec{a}$
axis) and to elongations along the $\eta$ and $\zeta$ directions (the orthorhombic $\vec{b}$ and $\vec{c}$
axes respectively).  In the following we consider an orthorhombic lattice with space group $Immm$, with
polymer
chains oriented along the $\vec{a}$ axis (Fig.\ 2).  We take the chains as rigid units of $D_{2h}$ symmetry where
the sole degree of freedom is the rotation angle $\psi$ about the chain axis $\vec{a}$.  The assumption of
a rigid chain is a reasonable first approximation, indeed vibrational density of states data on RbC$_{60}$
obtained by inelastic neutron scattering exhibit a low energy external mode below 5 meV which is interpreted
as arising from librations around the chain axis \cite{Sch}.  Inelastic neutron scattering results on the
orthorhombic phase of KC$_{60}$ are in close agreement with these results \cite{Gue}.

%-----------------------------------------------------------------------
%     figure 2
%-----------------------------------------------------------------------
\begin{figure} 
\resizebox{0.46\textwidth}{!}
{ 
 \includegraphics{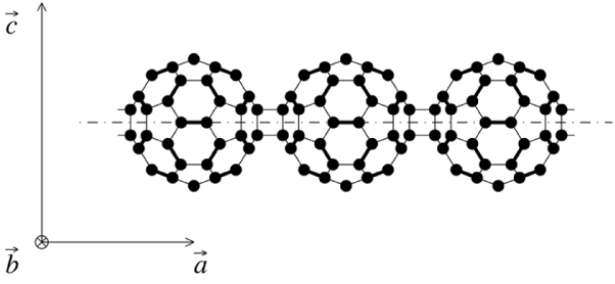} 
} 
\vspace{3mm}
%     figure caption
\caption{
Fragment of a C$_{60}$ polymer chain, orientation $\psi=0$.
} 
\label{fig2} 
\end{figure} 

The mathematical formulation of our model of a polymerized crystal is as follows.
We consider $N$ C$_{60}$ molecules located on a body-centered
orthorhombic lattice, where the center of mass positions of the molecules coincide with the lattice points.
The lattice points are labeled by indices $\vec{n}=(n_1,n_2,n_3)$.  These are either all integers
or all half integer numbers, corresponding respectively to the corners and the center of the orthorhombic
cell.
The equilibrium position of the $\vec{n}$-th lattice point then reads
\begin{equation}
\vec{X}(\vec{n})=n_1\vec{a}+n_2\vec{b}+n_3\vec{c},
\end{equation}
where $\vec{a}$, $\vec{b}$ and $\vec{c}$ are the orthorhombic lattice vectors.
The molecules (monomers) in a polymer chain are oriented so
that a twofold molecular axis coincides with the orthorhombic $\vec{a}$ axis ($\xi$-direction).  The number
of chains is $N_c$, and the number of monomers within a chain is $N_1$, hence $N=N_cN_1$.  We define as
standard orientation of a polymer chain
the orientation where the plane of cycloaddition is parallel to the $(\vec{a},\vec{c})$ planes of the
orthorhombic crystal.  The angle $\psi$ then measures a counterclockwise rotation of the polymer about
$\vec{a}$, the standard orientation corresponding to $\psi=0$ (Fig.\ 2).  Since it is sufficient to specify the
indices $\vec{\nu}=(n_2,n_3)$ to address a polymer chain, and because all molecules along one chain have the
same orientation, the rotation angle is independent of the index $n_1$ and is written as
$\psi\equiv\psi(\vec{\nu})$.
The interaction between chains is then formulated as a sum of two-body potentials
$U(\vec{n},\psi(\vec{\nu});\vec{n}',\psi(\vec{\nu}'))$ between C$_{60}$ monomers:
\begin{equation}
V=\frac{1}{2}\sum_{\vec{n},\vec{n}'}U(\vec{n},\psi(\vec{\nu});\vec{n}',\psi(\vec{\nu}')). \label{Vpotential}
\end{equation}
This interaction depends on the mutual orientation of the chains.  In order to
describe the symmetry reduction (phase transition) from a hypothetical polymer crystal with structure $Immm$
(i. e. no preferential orientation angle $\psi(\vec{\nu})$) to $Pmnn$ or $I2/m$, we perform a multipole
expansion of the potential in terms of SARFs $S_l(\vec{\nu})\equiv S_l(\psi(\vec{\nu}))=\sin(l\psi(\vec{\nu}))$,
$C_l(\vec{\nu})\equiv C_l(\psi(\vec{\nu}))=\cos(l\psi(\vec{\nu}))$, $l$ being the angular momentum quantum
number.
SARFs are the most appropriate variables for the description of orientation dependent properties in molecular
solids \cite{Jam,Pre,Yvi}.  Since in the present problem the assumption of a rigid chain leads to one single
rotation angle $\psi$, our SARFs are particularly simple and correspond to planar rotator functions
\cite{Pre2}.
Symmetry
of the chain for $\psi\longrightarrow \psi+\pi$ implies that only even values of $l$ occur.
In lowest order of the multipolar index $l$ we obtain
\begin{equation}
V=V_{QQ}^{RR}+V_{CF},
\end{equation}
with
\begin{mathletters}
\begin{eqnarray}
V_{QQ}^{RR} & = & \frac{1}{2}\sum_{n_1,n_1'}\sum_{\vec{\nu},\vec{\nu}'}J_{QQ}(n_1,\vec{\nu};n_1',\vec{\nu}')
S_2(\vec{\nu})S_2(\vec{\nu}'), \label{VQQRRformule}\\
V_{CF} & = & \sum_{n_1,n_1'}\sum_{\vec{\nu},\vec{\nu}'}v(n_1,\vec{\nu};n_1',\vec{\nu}')C_2(\vec{\nu}).
\end{eqnarray}
\end{mathletters}
Here $V_{QQ}^{RR}$ stands for the quadrupolar ($l=2$) rotation-rotation interaction
The quadrupolar function $S_2$, which is uneven in $\psi(\vec{\nu})$, is an appropriate order parameter.  The
coefficients $J_{QQ}$ are obtained from
\begin{eqnarray}
J_{QQ}(n_1,\vec{\nu};n_1',\vec{\nu}')=\frac{1}{\pi^2}
\int_{0}^{2\pi}d\psi(\vec{\nu})\int_{0}^{2\pi}d\psi(\vec{\nu}')  \nonumber   \\
\times\ U(\vec{n},\psi(\vec{\nu});\vec{n}',\psi(\vec{\nu}'))S_2(\psi(\vec{\nu}))S_2(\psi(\vec{\nu}')).
\end{eqnarray}
The term $V_{CF}$ accounts for the crystal field.  It is the potential experienced by a central chain with
orientation $\psi(\vec{\nu})$ where the surrounding chains are taken in cylindrical
approximation ($l=0$).  The coefficients $v$ are given by
\begin{eqnarray}
v(n_1,\vec{\nu};n_1',\vec{\nu}')=\frac{1}{2\pi^2}
\int_{0}^{2\pi}d\psi(\vec{\nu})\int_{0}^{2\pi}d\psi(\vec{\nu}') \nonumber \\
\times\ U(\vec{n},\psi(\vec{\nu});\vec{n}',\psi(\vec{\nu}'))C_2(\psi(\vec{\nu})). \label{v2coefficient}
\end{eqnarray}
For a C$_{60}$ monomer on a site $(n_1,\vec{\nu})$ we consider the interactions $J_a$ with the two monomers
at $(n_1\pm\frac{1}{2})$ on the four neighboring chains $(n_2\pm\frac{1}{2},n_3\pm\frac{1}{2})$ and the
interactions $J_b$ with the one monomer on each of the chains $(n_2\pm1,n_3)$.

We introduce Fourier transforms
\begin{mathletters}
\begin{eqnarray}
S_2(\vec{n})=\frac{1}{\sqrt{N}}\sum_{\vec{q}}S_2(\vec{q})e^{i\vec{q}\cdot\vec{X}(\vec{n})}, \\
S_2(\vec{q})=\frac{1}{\sqrt{N}}\sum_{\vec{n}}S_2(\vec{n})e^{-i\vec{q}\cdot\vec{X}(\vec{n})}.
\end{eqnarray}
\end{mathletters}
Since the chains are rigid and form a two-dimensional array $\vec{\nu}$,
$S_2(\vec{n})\equiv S_2(n_1,\vec{\nu})$ is independent of $n_1$, and hence $S_2(\vec{q})$ reduces to
\begin{equation}
S_2(0,\vec{q}_{\perp})=\sqrt{N_1}\delta_{q_{\xi}0}S_2(\vec{q}_{\perp}),
\end{equation}
with
\begin{equation}
S_2(\vec{q}_{\perp})=\frac{1}{\sqrt{N_c}}\sum_{\vec{\nu}}S_2(\vec{\nu})e^{-i\vec{q}_{\perp}\cdot
\vec{X}(\vec{\nu})}. \label{S2qloodrecht}
\end{equation}
Here we have $\vec{X}(\vec{\nu})=n_2\vec{b}+n_3\vec{c}$, and $\vec{q}_{\perp}=(q_{\eta},q_{\zeta})$, $q_{\xi}$, $q_{\eta}$, $q_{\zeta}$ being the components of
$\vec{q}$ along the orthorhombic axes.  The interaction $V^{RR}$ then reads
\begin{eqnarray}
V_{QQ}^{RR} & = & \frac{N_1}{2}\sum_{\vec{q}_{\perp}}J_{QQ}(\vec{q}_{\perp})S_2^\dagger(\vec{q_{\perp}})
S_2(\vec{q}_{\perp}), \label{VRR}\\
J_{QQ}(\vec{q}_{\perp}) & = & 8J_a\cos\left(\frac{q_{\eta}b}{2}\right)\cos\left(\frac{q_{\zeta}c}{2}\right)
+2J_b\cos\left(q_{\eta}b\right). \label{Jq}
\end{eqnarray}
In the following section we will derive explicit expressions for the quadrupolar interaction coefficients
$J_{a}$, $J_{b}$ and for the crystal field.  The crystal field for the same configuration of neighboring
interactions reads
\begin{equation}
V_{CF}=N_1\sum_{\vec{\nu}}\left(8v_a+2v_b\right)C_2(\vec{\nu}),
\end{equation}
where the coefficients $v_a$ and $v_b$ are determined by Eq.\ (\ref{v2coefficient}).  In the next section
the theory will be applied to the study of van der Waals and electrostatic quadrupole interactions between
C$_{60}$ polymer chains.
\end{section}
%
%
%--------------------------
%DIRECT INTERCHAIN COUPLING
%--------------------------
\begin{section}{Direct Interchain Coupling}\label{Section3}
The orientation dependent interaction between two polymer chains in the $Immm$ crystal of AC$_{60}$ is due to
direct van der Waals
interactions and to electric quadrupole forces between C$_{60}$ monomers.  The van der Waals interactions
also determine the mutual chain orientations in polymerized C$_{60}$.
%DIRECT VAN DER WAALS INTERACTIONS
%---------------------------------
\begin{subsection}{Direct van der Waals Interactions}
Here we consider repulsive Born-Mayer and attractive London dispersion forces between interaction centers
on monomers located in different chains.  The interaction centers comprise C atoms and double bond- and
single bond centers; they have been previously introduced for the description of the orientational disordered
and ordered phases of C$_{60}$ fullerite \cite{Spr,Hei,Lam2}.  While in orientationally disordered
C$_{60}$ all C atoms are equivalent, this is no longer the case in the polymer phase and we have to take
into account explicitly the geometrical constraints imposed by polymerization along the orthorhombic
$\vec{a}$ axis.  Throughout this paper we assume polymer chains of symmetry $D_{2h}$.  The neutron diffraction
results of Ref.\ \cite{Huq} show in fact that the symmetry of the chains in both KC$_{60}$ and RbC$_{60}$ is
lower, $C_{2h}$.  We attribute this reduction of symmetry to the molecular field of the originally ordered
structures $Pmnn$ and $I2/m$.  Since we start from an orientationally disordered structure $Immm$ and study
the transition towards $Pmnn$ or $I2/m$, we keep the chain symmetry $D_{2h}$.
It is convenient to classify the C atoms of a monomer in planar (100) sheets, perpendicular
to $\vec{a}$ (see Fig.\ 1).  We label these sheets by an index $\lambda=1,\ldots,17$.  The position of
sheet $\lambda$ with respect to the center of mass of the monomer is determined by a vector
$\vec{p}(\lambda)$:
\begin{equation}
\vec{p}(\lambda)=p(\lambda)\vec{e}_{\xi},
\end{equation}
with origin at the center of the monomer and end point at the intersection of sheet $\lambda$ with the
polymer axis.  This intersection is called the center of the sheet.  Here $\vec{e}_{\xi}$ is a unit vector
along the polymer axis.  The atoms within sheet $\lambda$ are labeled by an index $c(\lambda)$.  We will use
the notation $\Lambda=(\lambda,c(\lambda))$.  The
atomic positions relative to the center of mass of a monomer belonging to the polymer $\vec{\nu}$ which
is rotated away form the
standard orientation by the angle $\psi(\vec{\nu})$ then reads
\begin{eqnarray}
\vec{r}(\Lambda,\psi(\vec{\nu}))=r(\lambda)\{\cos[\phi(\Lambda)+\psi(\vec{\nu})]\vec{e}_{\eta} \nonumber \\
+\sin[\phi(\Lambda)+\psi(\vec{\nu})]\vec{e}_{\zeta}\} \label{rlambdaclambda}
\end{eqnarray}
with $r(\lambda)=d\sin\theta(\lambda)$.  Here $d$ is the radius of the C$_{60}$ molecule, $\theta(\lambda)$
and $\phi(\Lambda)$ are respectively the polar and the azimuthal angle of the
atom $c(\lambda)$ when the polymer is in the standard orientation.  The position of atom $\Lambda$ 
of the monomer at lattice site $\vec{n}=(n_1,\vec{\nu})$ is then given
by
\begin{equation}
\vec{R}(\vec{n},\Lambda,\psi(\vec{\nu}))=\vec{X}(\vec{n})+\vec{\rho}(\Lambda,\psi(\vec{\nu}))
\label{Rvector},
\end{equation}
where $\vec{\rho}(\Lambda,\psi(\vec{\nu}))=\vec{p}(\lambda)+\vec{r}(\Lambda,\psi(\vec{\nu}))$.
The distance between atoms $\Lambda$ and $\Lambda'$ belonging to monomers $\vec{n}$ and $\vec{n}'$ in
different chains $\vec{\nu}$ and $\vec{\nu}'$ reads
\begin{equation}
\Delta(\vec{n},\Lambda,\psi;\vec{n}',\Lambda',\psi')=|\vec{R}(\vec{n},\Lambda,\psi)
-\vec{R}(\vec{n}',\Lambda',\psi')|. \label{DELTA}
\end{equation}
Here we have written $\psi$, $\psi'$ for $\psi(\vec{\nu})$ and $\psi(\vec{\nu}')$ respectively.  The van der
Waals potential $U^W$ between these atoms is then given by
\begin{equation}
U^W(\vec{n},\Lambda,\psi;\vec{n}',\Lambda',\psi')=C_1\exp(-C_2\Delta)-B\Delta^{-6} \label{vdWpotential}
\end{equation}
with $\Delta$ given by Eq.\ (\ref{DELTA}).  Here $C_1$ and $C_2$ are the parameters of the repulsive
Born-Mayer
potential while $B$ determines the strength of the van der Waals attraction (London dispersion forces).  The
interaction potential between monomers then reads
\begin{equation}
U^W(\vec{n},\psi;\vec{n}',\psi')=\sum_{\Lambda,\Lambda'}U^W(\vec{n},\Lambda,\psi;\vec{n}',\Lambda',\psi')
\label{UW},
\end{equation}
where the double sum runs over the C atoms $\Lambda$, $\Lambda'$ of the two monomers.  In a similar way we can treat the
potential due to interaction centers located on double bonds and single bonds of the monomers.  For the case
of three interaction centers between two C atoms participating in a double bond on the monomer, the
present
formulation leads to the consideration of 26 additional planar sheets, the inclusion of single bond centers
adds 17 more sheets.  Details of the van~der~Waals potentials between various types of interaction centers
belonging to monomers on different chains are given in Appendix \ref{AppendixB}.  As a result the potential
between two monomers is again expressed by Eq.\ (\ref{UW}), where now the indices $\Lambda$ and $\Lambda'$ run
over atoms, double bond- and single bond interaction centers.  Having determined
$U^W(\vec{n},\psi;\vec{n}',\psi')$, we proceed as in Sect.\ \ref{rigid}, Eqs.\ 
(\ref{Vpotential})-(\ref{Jq}).
As a result we obtain the van der Waals contributions $J_a^W$ and $J_b^W$ to the quadrupolar
interactions $J_a$ and $J_b$.  
Numerical results are found in Table \ref{Table1}.  
%%%%%%%%%%%%%%%%%%%%%%%%%%%%%%%%%%%%%%%%%%%%%%%%%%%%%%%%%%%%%%%%%%%%%%%%%%%
%Table 1
\begin{table}
\caption{
Lattice constants of the cubic and orthorhombic lattices of AC$_{60}$ and C$_{60}$ (units {\AA}).  Calculated
direct interaction coefficients $J$ for van der Waals and Coulomb interactions (units Kelvin), orthorhombic
lattices.
\label{Table1}}
\begin{tabular}{ccccccccc}
& $a_c$ & $a_o$ & $b_o$ & $c_o$ & $J_a^W$ & $J_b^W$ & $J_a^C$ & $J_b^C$ \\
\tableline
KC$_{60}$ & $14.06$ & $9.11$ & $9.95$ & $14.32$ & $21.44$ & $-6.61$ & $10.70$ & $-55.17$ \\
%\tableline
RbC$_{60}$ & $14.08$ & $9.14$ & $10.11$ & $14.23$ & $21.33$ & $-2.88$ & $11.14$ & $-50.56$ \\
%\tableline
CsC$_{60}$ & $14.13$ & $9.10$ & $10.22$ & $14.17$ & $20.61$ & $-1.50$ & $11.65$ & $-47.68$ \\
%\tableline
C$_{60}$ & $14.15$ & $9.14$ & $9.90$ & $14.66$ & $13.13$ & $-8.41$ & / & /
\end{tabular}
\end{table}
%%%%%%%%%%%%%%%%%%%%%%%%%%%%%%%%%%%%%%%%%%%%%%%%%%%%%%%%%%%%%%%%%%%%%%%%%%%
There we quote the interchain van~der~Waals interaction
coefficients for orthorhombic lattice constants $a_o$, $b_o$, $c_o$ of polymerized AC$_{60}$, A=K, Rb, Cs and
polymerized C$_{60}$.  Similarly we determine the crystal field coefficients $v_a^W$ and $v_b^W$.
The results are quoted in Table \ref{Table2}.
%%%%%%%%%%%%%%%%%%%%%%%%%%%%%%%%%%%%%%%%%%%%%%%%%%%%%%%%%%%%%%%%%%%%%%%%%%%
%Table 2
\begin{table}
\caption{
Crystal field coefficients due to neighboring chains $v_a^W$ etc., units K.
\label{Table2}}
\begin{tabular}{ccccc}
& $v_a^W$ & $v_b^W$ & $v_a^C$ & $v_b^C$ \\
\tableline
KC$_{60}$ & $24.24$ & $-73.06$ & $-99.12$ & $422.53$ \\
RbC$_{60}$ & $23.44$ & $-15.31$ & $-92.99$ & $400.99$ \\
CsC$_{60}$ & $21.49$ & $5.57$ & $-89.55$ & $387.06$ \\
C$_{60}$ & $24.45$ & $-101.01$ & / & /
\end{tabular}
\end{table}
%%%%%%%%%%%%%%%%%%%%%%%%%%%%%%%%%%%%%%%%%%%%%%%%%%%%%%%%%%%%%%%%%%%%%%%%%%%
%
\end{subsection}
%
%
%Electrostatic Quadrupolar Interactions
%--------------------------------------
\begin{subsection}{Electrostatic Quadrupolar Interactions} \label{Sectionelectrostatic}
The charge transfer of one electron form the alkali atom to the C$_{60}$ molecule leads to an occupation of
the lowest unoccupied molecular orbital (LUMO) levels which are of $T_{1u}$ symmetry \cite{Had}.  Thereby the crystal
field of the C$_{60}^-$ ion acquires an electronic component that favors the same orientation of the
neighboring molecules along $[110]$ such that the stereospecific cycloaddition occurs \cite{Nik}.  We have
studied the electronic charge distribution on the C$_{60}^-$ units in the polymer chain by using a tight
binding model \cite{Nik2}.  The charge is mainly concentrated in the equatorial region of C$_{60}^-$, in
agreement with NMR results \cite{deS}.  We find that only even $l$ multipoles are allowed; in particular each C$_{60}$ unit has an electric quadrupole.  In the following
we adopt a simple model of charge distribution.  By using the labeling of C atoms of \cite{deS}, we locate a
charge of $-0.12$ (units electron charge $|e|$) on each bond C15-C16.  These charges are fixed at a
distance $d=3.52$ {\AA} from the center of the C$_{60}^-$ ball.  In the center we put a charge $-0.76$.  The
position of these three charges on a C$_{60}^-$ monomer centered at the lattice site
$\vec{n}=(n_1,\vec{\nu})$ belonging to chain $\vec{\nu}$ is then given by
\begin{equation}
\vec{R}(\vec{n},\alpha,\psi(\vec{\nu}))=\vec{X}(\vec{n})+\vec{D}(\alpha,\psi(\vec{\nu})),
\end{equation}
where
\begin{equation}
\vec{D}(\alpha,\psi(\vec{\nu}))=d(\alpha)\left[\vec{e}_{\eta}\cos\psi(\vec{\nu})
+\vec{e}_{\zeta}\sin\psi(\vec{\nu})\right],
\end{equation}
where $d(\alpha)=\pm d$ for $\alpha=1,2$ corresponding to the two charges $Q_{\alpha}=-0.12$ and $d(\alpha)=0$ for
$\alpha=3$, corresponding to the central charge $Q_{\alpha}=-0.76$.  The distance between two charges belonging to two monomers on
different chains is then given by
\begin{equation}
\Delta(\vec{n},\alpha,\psi;\vec{n}',\alpha',\psi')=|\vec{R}(\vec{n},\alpha,\psi)-\vec{R}(\vec{n}',\alpha',
\psi')|,
\end{equation}
where $\psi$, $\psi'$ stands for $\psi(\vec{}\nu)$, $\psi(\vec{\nu}')$.  The Coulomb interaction between
these two monomers then reads
\begin{equation}
U^C(\vec{n},\psi;\vec{n}',\psi')=\sum_{i,i'}F\frac{Q_{\alpha}Q_{\alpha'}}
{\Delta(\vec{n},\alpha,\psi;\vec{n}',\alpha',\psi')}.
\end{equation}
With our units of charges and with lengths in {\AA}, with $F=167000$ K {\AA}, the energy $U^C$ is measured
in units Kelvin.  One now proceeds as before with $U^W$.  We take into account the same configuration of
neighbors as before, sum over all chains in the crystal, expand the total potential in terms of SARFs and
obtain the interaction coefficients $J_a^C$, $J_b^C$ for the electrostatic quadrupole-quadrupole interaction
and the coefficients $v_a^C$ and $v_b^C$ for the electrostatic crystal field.  The results are also quoted in
Table \ref{Table1} and Table \ref{Table2} respectively.
Adding the quadrupole-quadrupole contributions from the van der Waals and the electrostatic interchain
potentials we obtain the total quadrupolar interaction $V_{QQ}^{RR}$, expression~(\ref{VRR}), where $J_{QQ}(
\vec{q}_{\perp})$ is given by Eq.\ (\ref{Jq}) with
\begin{mathletters}
\begin{eqnarray}
J_a=J_a^W+J_a^C, \label{Jaformule}\\
J_b=J_b^W+J_b^C. \label{Jbformule}
\end{eqnarray}
\end{mathletters}

Notice that for polymerized pristine C$_{60}$, there is no electrostatic quadrupole due to charge transfer,
and hence $J_a^C$ and $J_b^C$ are zero.  We observe that in principle the deformation of the monomer in
C$_{60}$ from $I_h\longrightarrow D_{2h}$ leads to a redistribution of the electric charge on the neutral
molecule.  Here it is assumed that these effects are included in $J_a^W$ and $J_b^W$.

We return to the quadrupolar interaction $J(\vec{q})$ and study its wave number dependence.  In
two-dimensional $\vec{q}_{\perp}$ space $(q_{\eta},q_{\zeta})$ we have at the Brillouin zone (BZ) center
$\vec{q}_{\perp}=(0,0)=\vec{q}_{\Gamma}$
\begin{equation}
J_{QQ}(\vec{q}_{\Gamma})=8J_a+2J_b. \label{JQQqgammaformule}
\end{equation}
With the numerical values of Table \ref{Table1} we conclude that for polymerized AC$_{60}$, A=K, Rb, Cs as
well as for polymerized C$_{60}$, $J(\vec{q}_{\Gamma})>0$.  On the other hand, at the Brillouin zone boundary
$\vec{q}_{\perp}=(0,\frac{2\pi}{c})=\vec{q}_Z$ we get
\begin{equation}
J_{QQ}(\vec{q}_Z)=-8J_a+2J_b,
\end{equation}
with $J(\vec{q}_Z)<0$ and $|J(\vec{q}_Z)|>J(\vec{q}_\Gamma)$.  The dominance and negative sign of
$J(\vec{q}_Z)$ leads to a condensation of $S_2(\vec{q})$ at $\vec{q}_{\perp}=\vec{q}_Z$:
\begin{equation}
\left<S_2(\vec{q}_{\perp})\right>=\sqrt{N_c}\sigma\delta_{\vec{q}_{\perp},\vec{q}_Z},
\end{equation}
which is the condensation scheme proposed earlier \cite{Aks}.
Here $\left<\;\right>$ stands for a thermal average over the crystal, $\sigma$ is the order parameter
amplitude.
The number of chains $N_c$ corresponds to the number of lattice points in the orthorhombic
$(\vec{b},\vec{c})$
plane.  Condensation at $\vec{q}_Z$ implies that the chains in the same basal plane $(\vec{a},\vec{b})$ of
the
orthorhombic lattice all have the same orientation, but the orientation alternates in neighboring planes at
distance $c/2$.  This is the ``antiferrorotational" structure $Pmnn$ (Fig.\ 1a).  We find that the
rotation-rotation interaction due to van der Waals forces and electrostatic quadrupolar forces leads to
$Pmnn$ for
KC$_{60}$, RbC$_{60}$ and CsC$_{60}$, irrespective of the orthorhombic lattice constants.  While for
KC$_{60}$ this structure has been found by experiment \cite{Ste,Lau}, the experimental result for RbC$_{60}$
\cite{Lau,Huq} and
CsC$_{60}$ \cite{Rou} is $I2/m$.  We conclude that the alkalis must play a specific role and we will
investigate this problem in the next section.

On the other hand, for polymerized pristine C$_{60}$, our analysis confirms the experimental result $Pmnn$
\cite{Lau2,Dav2}.
\end{subsection}
\end{section}
%
%
%----------------------------
%Alkali-mediated Interactions
%----------------------------
\begin{section}{Alkali-mediated Interactions}\label{Sectionalkalimediated}
% Translation-Rotation Coupling
%------------------------------
\begin{subsection}{Translation-Rotation Coupling}
In orientationally disordered molecular crystals or in crystals containing ions with orientational disorder,
the coupling of molecular rotations with center of mass lattice displacements of molecules and of surrounding
ions plays an important role in determining the structural properties.  For a review on the so-called
translation-rotation (TR) coupling, see Ref.\ \cite{Lyn}.  Here we will study the corresponding coupling of
the rotational motion of polymer chains with the lattice displacements of the surrounding alkali ions.  We
extend the model of an orthorhombic crystal consisting of orientationally disordered rigid polymers
with axis of polymerization along the orthorhombic $\xi$ direction (Sect. \ref{rigid}) by including the
alkali metal ions at equilibrium lattice positions
\begin{equation}
\vec{X}(\vec{n}_A)=n_{A1}\vec{a}+n_{A2}\vec{b}+n_{A3}\vec{c},
\end{equation}
where $\vec{n}_A=(n_{A1},n_{A2},n_{A3})$ are the lattice indices of the alkali atoms.  For a
C$_{60}^-$ monomer at lattice site $\vec{n}=(n_1,n_2,n_3)$, the six nearest neighbor alkalis are located
at $(n_1\pm\frac{1}{2},n_2\pm\frac{1}{2},n_3)$, $(n_1,n_2,n_3\pm\frac{1}{2})$.  Taking into account lattice
displacements $\vec{u}_A(\vec{n}_A)$ we write for the actual positions of the alkalis
\begin{equation}
\vec{R}(\vec{n}_A)=\vec{X}(\vec{n}_A)+\vec{u}_A(\vec{n}_A).
\end{equation}
%where $\vec{\tau}_A=\vec{X}(\vec{n}_A)-\vec{X}(\vec{n})$.
The distance between a C atom in position $\Lambda$ belonging to a
C$_{60}^-$ monomer at lattice site $\vec{n}=(n_1,\vec{\nu})$ in the polymer chain $\vec{\nu}$ and a
surrounding alkali ion at $\vec{n}_A$ is given by $|\vec{R}(\vec{n},\Lambda,\psi(\vec{\nu}))-
\vec{R}(\vec{n}_A)|$ or equivalently by
\begin{equation}
\Delta(\vec{n},\Lambda,\psi(\vec{\nu});\vec{n}_A)=
|\vec{\rho}(\Lambda,\psi(\vec{\nu}))+\vec{X}(\vec{n}-\vec{n}_A)-\vec{u}_A(\vec{n}_A)|.
\end{equation}
Here we have used expression (\ref{Rvector}) for $\vec{R}(\vec{n},\Lambda,\psi(\vec{\nu}))$,
$\psi(\vec{\nu})$
being the orientation angle of the polymer chain.  The van der Waals interaction of the monomer and the
alkali ion is then given by
\begin{equation}
U^W(\vec{n},\psi(\vec{\nu});\vec{n}_A)=\sum_{\Lambda}C_1\exp(-C_2\Delta)-B\Delta^{-6},
\end{equation}
where the potential parameters $C_1$, $C_2$ and $B$ refer to the C-A interaction (see Appendix \ref{AppendixB}).
The Coulomb interaction potential of the quadrupolar electric charges $Q_{\alpha}$ on the C$_{60}^-$ monomers
(See Sect. \ref{Sectionelectrostatic}) with an alkali ion A$^+$ with unit charge $Q^A$ reads:
\begin{mathletters}
\begin{equation}
U^C(\vec{n},\psi(\vec{\nu});\vec{n}_A)=F\sum_{\alpha}\frac{Q_{\alpha}Q^A}
{\Delta(\vec{n},\alpha,\Lambda,\psi(\vec{\nu});\vec{n}_A)}
\end{equation}
where
\begin{eqnarray}
& & \Delta(\vec{n},\alpha,\Lambda,\psi(\vec{\nu});\vec{n}_A) \nonumber \\
& & =|\vec{D}(i,\psi(\vec{\nu}))+\vec{X}(\vec{n}-\vec{n}_A)
-\vec{u}_A(\vec{n}_A)|
\end{eqnarray}
\end{mathletters}
is the distance between the charge $\alpha$ on the monomer and the center of the alkali ion.  Assuming
that the lattice displacements are small (in comparison to the lattice constants), we expand these
potentials ($U$ stands for $U^W$ or $U^C$) in terms of $\vec{u}_A$ and retain the first two terms:
\begin{eqnarray}
U(\vec{n},\psi(\vec{\nu});\vec{n}_A)=U^{(0)}(\vec{n}-\vec{n}_A,\psi(\vec{\nu})) \nonumber \\
+\sum_{i}U_i^{(1)}(\vec{n}-\vec{n}_A,\psi(\vec{\nu}))u_{Ai}(\vec{n}_A), \label{Uexpansion}
\end{eqnarray}
where $i$ labels the Cartesian components along the orthorhombic axes.
The first term on the right hand side (RHS) corresponds to
the situation $\vec{u}_A=\vec{0}$, the alkali ion is at its equilibrium position.  Exploiting the angular
dependence of $U^{(0)}$ we expand it in terms of SARFs.  Since each monomer is located at a center of
symmetry
with respect to the surrounding alkalis in the orthorhombic lattice, only $C_l(\psi(\vec{\nu}))$ terms
contribute to
\begin{equation}
\sum_{\vec{n}_A}U^{(0)}(\vec{n}-\vec{n}_A,\psi(\vec{\nu}))=
\sum_{\vec{n}_A}v^{(0)}(\vec{n}_A-\vec{n})C_2(\psi(\vec{\nu})) \label{dubbelesom},
\end{equation}
where
\begin{equation}
v^{(0)}(\vec{n}_A-\vec{n})=\frac{1}{\pi}\int_0^{2\pi}d\psi(\vec{\nu})
U^{(0)}(\vec{n}-\vec{n}_A,\psi(\vec{\nu}))C_2(\psi(\vec{\nu})).
\end{equation}
Here we restrict ourselves to the lowest order multipoles with $l=2$.  Expression (\ref{dubbelesom}) 
stands for the interaction of one monomer at site $n_1$ in chain $\vec{\nu}$ with six surrounding alkali
atoms at equilibrium lattice sites $\vec{X}(\vec{n}_A)$.  Since all monomers of the chain are equivalent,
the RHS result does not depend on $n_1$.  Taking into account $U^C$ and $U^W$ and summing over all monomer
sites, we obtain the crystal field potential due to the nearest neighbor alkali atoms
\begin{equation}
V^{A}_{CF}=N_1\sum_{\vec{\nu}}(4v^A_{ab}+2v^A_c)C_2(\psi(\vec{\nu}))
\end{equation}
where we have
\begin{mathletters}
\begin{eqnarray}
v^A_{ab} & = & v_{ab}^{A,W}+v_{ab}^{A,C}, \\
v^A_c & = & v_c^{A,W}+v_c^{A,C}.
\end{eqnarray}
\end{mathletters}
Here $v_{ab}$ and $v_c$ refer to alkalis located at $(\pm\frac{1}{2},\pm\frac{1}{2},0)$ and
$(0,0,\pm\frac{1}{2})$ respectively, counted from the monomer center $\vec{n}$.  Numerical values of these
coefficients are quoted in Table \ref{Table3}.
%%%%%%%%%%%%%%%%%%%%%%%%%%%%%%%%%%%%%%%%%%%%%%%%%%%%%%%%%%%%%%%%%%%%%%%%%%%
%Table 3
\begin{table}
\caption{
Crystal field coefficients due to alkalis, units K.
\label{Table3}}
\begin{tabular}{ccccc}
& $v_{ab}^{A,W}$ & $v_c^{A,W}$ & $v_{ab}^{A,C}$ & $v_c^{A,C}$ \\
\tableline
KC$_{60}$ & $-10.55$ & $16.38$ & $-497.98$ & $1137.42$ \\
RbC$_{60}$ & $-14.30$ & $26.44$ & $-493.28$ & $1161.03$ \\
CsC$_{60}$ & $-19.91$ & $30.44$ & $-497.81$ & $1177.16$
\end{tabular}
\end{table}
%%%%%%%%%%%%%%%%%%%%%%%%%%%%%%%%%%%%%%%%%%%%%%%%%%%%%%%%%%%%%%%%%%%%%%%%%%%

The second term on the RHS of Eq.\ (\ref{Uexpansion}) is the proper TR contribution, $U_i^{(1)}$ stands for
the first order derivative of $U$ with respect to $u_{Ai}$.  Expanding $U_i^{(1)}$ in
terms of SARFs, summing over the surrounding six alkalis of the monomer and exploiting the symmetry of the
orthorhombic lattice, we find in lowest order $l=2$ of the
multipole expansion:
\begin{eqnarray}
& & \sum_{\vec{n}_A}U_i^{(1)}(\vec{n}-\vec{n}_A,\psi(\vec{\nu}))u_{Ai}(\vec{n}_A) \nonumber \\
& & =
  \sum_{\vec{n}_A}v_i^{(1)}(\vec{n}_A-\vec{n})S_2(\psi(\vec{\nu}))u_{Ai}(\vec{n}_A).
\end{eqnarray}
Here we retain only the terms containing the function $S_2$ and drop those with $C_2$.  Indeed it can be
shown that TR terms with $C_2$ do not contribute to a change of the orthorhombic lattice structure.  The
coefficients $v_i^{(1)}(\vec{n}_A-\vec{n})$ are obtained by
\begin{equation}
v_i^{(1)}(\vec{n}_A-\vec{n})=\frac{1}{\pi}\int_0^{2\pi}d\psi(\vec{\nu})U_i^{(1)}(\vec{n}-\vec{n}_A,\psi(\vec{\nu}))S_2(\psi(\vec{\nu})) \label{vitauA}.
\end{equation}
Summing over all monomers in the crystal, the total TR potential due to $U^W+U^C$ is found to be
\begin{equation}
V^{TR}=\sum_{\vec{n}}\sum_{\vec{n}_A}\sum_i v_i^{(1)}(\vec{n}_A-\vec{n})S_2(\psi(\vec{\nu}))
u_{Ai}(\vec{n}_A). \label{UTRsumoverallmonomers}
\end{equation}
We mention the symmetry property
\begin{equation}
v_i^{(1)}(\vec{n}_A-\vec{n})=-v_i^{(1)}(\vec{n}-\vec{n}_A).
\end{equation}
From expression (\ref{vitauA}) it follows that all six coefficients $v_1^{(1)}(\vec{n}_A-\vec{n})$ are zero.
The only nonzero coefficients are $v_2^{(1)}(0,0,\pm\frac{1}{2})=\mp v^A_{2,c}$,
$v_3^{(1)}(\pm\frac{1}{2},\pm\frac{1}{2},0)=\mp v^A_{3,ab}$ and
$v_3^{(1)}(\pm\frac{1}{2},\mp\frac{1}{2},0)=\mp v^A_{3,ab}$, where
\begin{mathletters}
\begin{eqnarray}
v^A_{2,c} & = & v_{2,c}^{A,W}+v_{2,c}^{A,C}, \\
v^A_{3,ab} & = & v_{3,ab}^{A,W}+v_{3,ab}^{A,C}.
\end{eqnarray}
\end{mathletters}
Numerical values are given in Table \ref{Table4}.
%%%%%%%%%%%%%%%%%%%%%%%%%%%%%%%%%%%%%%%%%%%%%%%%%%%%%%%%%%%%%%%%%%%%%%%%%%%
%Table 4
\begin{table}
\caption{
Calculated indirect interactions coefficients for TR coupling ($v$ in units K/{\AA}).  Electronic
polarizability parameters, $d_A$ in units {\AA}, $g_A$ and calculated $\lambda_b$, $\lambda_c$ in units K.
\label{Table4}}
\begin{tabular}{ccccccccc}
& $v_{2,c}^{A,W}$ & $v_{3,ab}^{A,W}$ & $v_{2,c}^{A,C}$ & $v_{3,ab}^{A,C}$ & $d_A$ & $\lambda_b$ & $\lambda_c$ & $g_A$ \\
\tableline
KC$_{60}$ & $4.57$ & $4.24$ & $317.71$ & $199.98$ & $1.47$ & $9.28$ & $51.20$ & $43.25$ \\
%\tableline
RbC$_{60}$ & $7.48$ & $5.70$ & $326.36$ & $195.17$ & $1.82$ & $13.25$ & $83.52$ & $35.00$ \\
%\tableline
CsC$_{60}$ & $8.59$ & $7.79$ & $332.29$ & $194.84$ & $1.87$ & $14.27$ & $90.75$ & $34.00$ \\
\end{tabular}
\end{table}
%%%%%%%%%%%%%%%%%%%%%%%%%%%%%%%%%%%%%%%%%%%%%%%%%%%%%%%%%%%%%%%%%%%%%%%%%%%
We define Fourier transforms of displacements by
\begin{mathletters}
\begin{eqnarray}
\vec{u}_A(\vec{n}_A) & = & \frac{1}{\sqrt{Nm_A}}\sum_{\vec{q}}e^{i\vec{q}\cdot\vec{X}(\vec{n}_A)}
\vec{u}_A(\vec{q}), \\
\vec{u}_A(\vec{q}) & = & \sqrt{\frac{m_A}{N}}\sum_{\vec{n}_A}e^{-i\vec{q}\cdot\vec{X}(\vec{n}_A)}
\vec{u}_A(\vec{n}_A).
\end{eqnarray}
\end{mathletters}
Using in addition Eq.~(\ref{S2qloodrecht}), we rewrite Eq.~(\ref{UTRsumoverallmonomers}):
\begin{equation}
V^{TR}=\sqrt{\frac{N_1}{m_A}}\sum_{\vec{q}_{\perp}}\sum_i v_i^{(1)}(\vec{q}_{\perp})S_2^{\dagger}
(\vec{q}_{\perp})
u_{Ai}(0,\vec{q}_{\perp}), \label{VTRformule}
\end{equation}
where
\begin{equation}
v_i^{(1)}(\vec{q}_{\perp})=i\sum_{\vec{n}_A}v^{(1)}_i(\vec{n}_A-\vec{n})\sin(\vec{q}\cdot
\vec{X}(\vec{n}_A-\vec{n})),
\end{equation}
or equivalently
\begin{equation}
\vec{v}^{(1)}(\vec{q}_{\perp})=\left(\begin{array}{c}
0 \\
-2iv_{2,c}^A\sin(q_{\zeta}\frac{c}{2}) \\
-4iv_{3,ab}^A\sin(q_{\eta}\frac{b}{2}) \end{array}\right).
\end{equation}
We see that the rotational motion of the chains about the $\vec{a}$ axis induces displacements of the alkali
atoms only along the $\vec{b}$ and $\vec{c}$ axes.  We observe that the coupling $v_i^{(1)}(\vec{q}_{\perp})$
vanishes at $\vec{q}_{\perp}=\vec{q}_Z$.

In order to relate our results to elastic deformations of the lattice, we consider the RHS of Eq.\
(\ref{VTRformule}) in the long wavelength regime and transform to acoustic (ac) lattice displacements
\begin{equation}
\vec{s}(\vec{q})=\sqrt{\frac{m_A}{m}}\vec{u}_A(\vec{q})+\sqrt{\frac{m_{C_{60}}}{m}}\vec{u}_{C_{60}}(\vec{q}),
\end{equation}
where $m=m_A+m_{C_{60}}$ is the total mass per primitive unit cell, $m_{C_{60}}$ being the mass of a C$_{60}$
monomer.  The acoustic part of $V^{TR}$ then reads
\begin{equation}
V_{ac}^{TR}=\sqrt{N_1}\sum_{\vec{q}_{\perp}}\sum_i \hat{v}_i^{(1)}(\vec{q}_{\perp})
s_i(0,\vec{q}_{\perp})S_2^{\dagger}(\vec{q}_{\perp}).
\end{equation}
The sum over $i$ is now restricted to the components $\eta$ and $\zeta$ and hence
\begin{equation}
\vec{\hat{v}}_{ac}^{(1)}(\vec{q}_{\perp})=-\frac{i}{\sqrt{m}}\left(\begin{array}{c}
cq_{\zeta}v_{2,c}^A \\
2bq_{\eta}v_{3,ab}^A \end{array}\right).
\end{equation}
The translational part of the acoustic lattice energy is given by
\begin{equation}
V_{ac}^{TT}=\frac{1}{2}\sum_{\vec{q}}\sum_{i,j}s_i^{\dagger}(\vec{q})M_{ij}(\vec{q})s_j(\vec{q}).
\end{equation}
Here $M_{ij}(\vec{q})$ is the dynamical matrix of the orthorhombic crystal in absence of TR coupling.  In the
following we need only to retain the components $(\eta,\zeta)$ of $\vec{s}(0,\vec{q_{\perp}})$.  Then
$M(\vec{q})$ reduces to a $2\times2$ matrix:
\begin{equation}
M(\vec{q}_{\perp})=\frac{1}{\rho}\left(\begin{array}{cc}
q_{\eta}^2c_{22}+q_{\zeta}^2c_{44} & q_{\eta}q_{\zeta}(c_{23}+c_{44}) \\
q_{\eta}q_{\zeta}(c_{23}+c_{44}) & q_{\zeta}^2c_{33}+q_{\eta}^2c_{44} \end{array}\right).
\end{equation}
Here $\rho$ is the mass density in the body centered orthorhombic unit cell, $\rho=2m/(abc)$.  We use the
Voigt-notation for the orthorhombic elastic constants $c_{22}$ etc..  We now consider
\begin{equation}
V_{ac}=V_{ac}^{TR}+V_{ac}^{TT}.
\end{equation}
For a fixed configuration of orientations $\{S_2(\vec{q}_{\perp})\}$, we minimize $V_{ac}$ with respect to
$s_i(0,\vec{q}_{\perp})$, $i=\eta,\zeta$, and obtain
\begin{equation}
s^{\dagger}(0,\vec{q}_{\perp})=-\sqrt{N_1}M^{-1}(\vec{q}_{\perp})\hat{v}^{(1)}(\vec{q}_{\perp})S_2^{\dagger}
(\vec{q}_{\perp}). \label{sdagger0q}
\end{equation}
Substitution of this expression into $V_{ac}$ leads to the effective rotation-rotation interaction which we
denote by $V_{ac}^{RR}$:
\begin{equation}
V_{ac}^{RR}=-\frac{N_1}{2}\sum_{\vec{q}_{\perp}}S_2^{\dagger}(\vec{q}_{\perp})J_{ac}(\vec{q}_{\perp})
S_2(\vec{q}_{\perp}), \label{VacRR}
\end{equation}
where
\begin{equation}
J_{ac}(\vec{q}_{\perp})=\vec{\hat{v}}_{ac}^{(1)\dagger}(\vec{q}_{\perp})M^{-1}(\vec{q}_{\perp})
\hat{v}_{ac}^{(1)}(\vec{q}_{\perp}). \label{Jacformule}
\end{equation}
Since $J_{ac}(\vec{q}_{\perp})>0$, the lattice mediated interaction $V_{ac}^{RR}$ is always attractive.
The largest value is obtained for $\vec{q}_{\perp}=\vec{q}_{\Gamma}=\vec{0}$:
\begin{equation}
\lim_{q_{\zeta}\longrightarrow 0}\left.J_{ac}(\vec{q}_{\perp})\right|_{q_{\eta}=0}=
\frac{2(v_{2,c}^A)^2c}{c_{44}ab}\equiv J_{ac}(\vec{q}_{\Gamma}). \label{Jacqgammaformule}
\end{equation}
With $c_{44}=870$ K \AA$^{-3}$ taken from Ref.\ \cite{YuB} for C$_{60}$-fullerite, and the values
of $v_{2,c}^A$ from Table \ref{Table4}, we find $J_{ac}(\vec{q}_{\Gamma})$ as quoted in Table \ref{Table5}
for KC$_{60}$, RbC$_{60}$ and CsC$_{60}$.
%%%%%%%%%%%%%%%%%%%%%%%%%%%%%%%%%%%%%%%%%%%%%%%%%%%%%%%%%%%%%%%%%%%%%%%%%%%
%Table 5
\begin{table}
\caption{
Interactions at $\vec{q}_{\perp}=\vec{q}_{\Gamma}$, units K.
\label{Table5}}
\begin{tabular}{cccccc}
& $J_{QQ}^C(\vec{q}_{\Gamma})$ & $J_{QQ}^W(\vec{q}_{\Gamma})$ & $-J_{ac}(\vec{q}_{\Gamma})$ & $-J_{ep}(\vec{q}_{\Gamma})$ & $J(\vec{q}_{\Gamma})$ \\
\tableline
KC$_{60}$ & $-24.80$ & $158.30$ & $-37.72$ & $-225.02$ & $-129.24$ \\
%\tableline
RbC$_{60}$ & $-12.03$ & $164.88$ & $-39.44$ & $-691.62$ & $-578.21$ \\
%\tableline
CsC$_{60}$ & $-2.18$ & $161.88$ & $-40.65$ & $-837.14$ & $-718.09$ \\
%\tableline
C$_{60}$ & / & $88.22$ & / & / & $88.22$ 
\end{tabular}
\end{table}
%%%%%%%%%%%%%%%%%%%%%%%%%%%%%%%%%%%%%%%%%%%%%%%%%%%%%%%%%%%%%%%%%%%%%%%%%%%
The attractive interaction at $\vec{q}_{\Gamma}$ favors a condensation of a ferrorotational structure
where all polymer chains have the same orientation, i.~e.~the space group $I2/m$.  However comparison of the
numerical values of $J_{ac}(\vec{q}_{\Gamma})$ and $J(\vec{q}_Z)$ shows that the interaction, which leads to
the antiferrorotational structure $Pmnn$, is dominant.  Hence the TR coupling mechanism is not sufficient to
explain the structural difference between KC$_{60}$ and Cs-, RbC$_{60}$.  In part \ref{partB} we will exploit
a different alkali-mediated interaction mechanism.  It is based on a specific {\em quadrupolar} electronic
polarizability of the alkali ions.
\end{subsection}
%
%
%Quadrupolar Polarizability
%--------------------------
\begin{subsection}{Quadrupolar Polarizability}\label{partB}
We will now include the role of the specific electronic polarizability of the alkali ions.  It is known from
work on the ammonium halides \cite{Hul} that the indirect interaction of two NH$_4^+$ tetrahedra via the
polarizable halide ions plays an essential role in determining the various crystalline phases of the ammonium
halides NH$_4$X, X = Cl, Br and I.  However, in the present problem, since the C$_{60}$ monomers
have symmetry $D_{2h}$, they do not couple to the dipolar electronic polarizability of the alkali metal ions.
We have to resort to the quadrupolar polarizability.  Since the C$_{60}^-$ units in a polymer chain are
rigidly linked in the same orientation, the C$_{60}^-$ chains, which support electric quadrupoles, produce
coherent electric field gradients which induce an anisotropic (quadrupolar) deformation of the residual
electronic charge of the alkalis.
The presence of residual electronic charge is seen as a consequence of the uniquely large interstitial space
between the C$_{60}$ molecules (we recall that in AC$_{60}$, the A$^+$ cations occupy the formerly octahedral
positions of the cubic phase).  Within a touching sphere model, with 5 {\AA} as the effective van der Waals
radius of the C$_{60}$ molecule \cite{Kra} and a cubic lattice constant $a=14.15$ {\AA}, we estimate the
radius of the interstitial sphere to be $R_A\approx2.075$ {\AA}.  Although the charge transfer from the alkali
atom to the C$_{60}$ molecule is considered as complete, there still will be a charge of $0.1-0.15|e|$ left
inside the
interstitial sphere centered at each alkali site.  In Appendix \ref{AppendixC} we discuss in more detail the
microscopic
origin of the quadrupolar polarizability of the cations.  Within a tight-binding model of the conduction
electron band, we expand the electron wave function at an alkali site in terms of local $s$ and $d$
functions.  If there is an appreciable weight of $d$-states, then the alkalis acquire a quadrupolar moment.
We model the corresponding charge distribution of each alkali ion by a symmetric linear dumbbell centered
on lines along the $\vec{a}$ direction.  We take dumbbells with equal charges
$Q_{\beta}=q_A$, $\beta=1,2$, at distances $\pm d_A$ from the center.  The numerical values of $d_A$ (Table \ref{Table4}) are the
average radii of valence electron $d$ shells calculated with atomic wave functions $3d_{3/2}$, $4d_{3/2}$ and
$5d_{3/2}$ for K$^+$, Rb$^+$ and Cs$^+$ respectively (Appendix \ref{AppendixC}).  We observe that the $d_A$ values for Cs$^+$ and
Rb$^+$ are close to each other but differ from K$^+$.  We consider $d$ shells because they can support an
electric quadrupole moment.  On a same crystalline line along $\vec{a}$, these dumbbells are
parallel with their axis perpendicular to $\vec{a}$ and a same orientation angle $\psi$ with the $\vec{c}$
axis.  In the crystal we then have chains of alkali dumbbells, where the rigid chain is not imposed by
intrachain interactions (as is the case for the C$_{60}$ polymers formed by cycloaddition) but by the
surrounding C$_{60}^-$ chains.  The location of a dumbbell in the crystal is determined by $\vec{n}_A=(n_{A1}
,\vec{\nu}_A)$, where $\vec{\nu}_A=(n_{2A},n_{3A})$ denotes the chain and where $n_{A1}$ labels the dumbbell
within the chain.  The orientational motion of an alkali dumbbell is characterized by SARFs $s_2(\vec{\nu}_A)
=\sin(2\psi(\vec{\nu}_A))$, independent of $n_{A1}$.  The electric quadrupole-quadrupole interaction
potential between a C$_{60}$ monomer at $\vec{n}$ and a surrounding alkali dumbbell at $\vec{n}_A$ is given
by
\begin{eqnarray}
& & U^{QQ}(\vec{n},\psi(\vec{\nu});\vec{n}_A,\psi(\vec{\nu}_A)) \nonumber \\
& & =
F\sum_{\alpha,\beta}\frac{Q_{\alpha}Q_{\beta}}{\Delta(\vec{n},\alpha,\psi(\vec{\nu});\vec{n}_A,\beta,
\psi(\vec{\nu}_A))} \label{UQQpotential}
\end{eqnarray}
where
\begin{eqnarray}
& & \Delta(\vec{n},\alpha,\psi(\vec{\nu});\vec{n}_A,\beta,\psi(\vec{\nu}_A)) \nonumber \\
& & =
|\vec{D}(\alpha,\psi(\vec{\nu}))+\vec{X}(\vec{n}-\vec{n}_A)-\vec{d}_A(\beta,\psi(\vec{\nu}_A))|
\end{eqnarray}
is the distance between a charge $Q_{\alpha}$ on the monomer and a charge $Q_{\beta}$ on the alkali dumbbell.
The position of the charge $\beta$ on the dumbbell $\vec{n}_A$ in the crystal reads
\begin{equation}
\vec{X}(\vec{n}_A,\beta)=\vec{X}(\vec{n}_A)+\vec{d}_A(\beta,\psi(\vec{\nu}_A)),
\end{equation}
where $\vec{d}_A$ is the position vector of the charge with respect to the center of the dumbbell.
Explicitly we have
\begin{equation}
\vec{d}_A(\beta,\psi(\vec{\nu}_A))=\pm d_A\left\{\cos[\psi(\vec{\nu}_A)]\vec{e}_{\eta}+
\sin[\psi(\vec{\nu}_A)]\vec{e}_{\zeta}\right\},
\end{equation}
where $+$ and $-$ refer to the two quadrupolar charges.

The potential (\ref{UQQpotential}) is expanded in SARFs.  After summation over the polymer chains $\vec{n}$
and the surrounding alkalis $\vec{n}_A$ we find (compare with Eq.\ (\ref{VQQRRformule})):
\begin{equation}
V^{Ss}=\sum_{n_1,n_{A1}}\sum_{\vec{\nu},\vec{\nu}_A}\lambda(n_1,\vec{\nu};n_{A1},\vec{\nu}_A)S_2(\vec{\nu})
s_2(\vec{\nu}_A),
\end{equation}
where
\begin{eqnarray}
\lambda(n_1,\vec{\nu};n_{A1},\vec{\nu}_A)=\frac{1}{\pi^2}\int_0^{2\pi}d\psi(\vec{\nu})
\int_0^{2\pi}d\psi(\vec{\nu}_A) \nonumber \\
\times\ U^{QQ}(\vec{n},\psi(\vec{\nu});\vec{n}_A,\psi(\vec{\nu}_A))S_2(\psi(\vec{\nu}))
s_2(\vec{\nu}_A)).
\end{eqnarray}
Going over to Fourier space we have
\begin{equation}
V^{Ss}=N_1\sum_{\vec{q}_{\perp}}\lambda(\vec{q}_{\perp})S_2^{\dagger}(\vec{q}_{\perp})s_2(\vec{q}_{\perp}),
\end{equation}
where
\begin{equation}
\lambda(\vec{q}_{\perp})=4\lambda_b\cos\left(\frac{q_{\eta}b}{2}\right)
+2\lambda_c\cos\left(\frac{q_{\zeta}c}{2}\right) \label{lambdaqloodrechtformule}.
\end{equation}
Here we have restricted ourselves to the six nearest neighbors $\vec{n}_A$ of a given monomer $\vec{n}$.
With the values $d_A$ of Table \ref{Table4} and with the same value $q_A=0.12$ for the three compounds,
we have calculated the interaction energies $\lambda_b$ and $\lambda_c$ quoted in Table \ref{Table4}.
The quantity $|\lambda(\vec{q}_{\perp})|$ is maximum at $\vec{q}_{\perp}=\vec{q}_{\Gamma}$ in
contradistinction with the direct RR interaction $J_{QQ}(\vec{q}_{\perp})$, Eq.~(\ref{Jq}).  The intraionic
restoring
forces of the electronic shells of the cations are described by a sum of single particle energy terms
\begin{equation}
V^{ss}=g_A\sum_{\vec{n}_A}s_2^2(\vec{n}_A)=N_1g_A\sum_{\vec{q}_{\perp}}s_2^{\dagger}(\vec{q}_{\perp})
s_2(\vec{q}_{\perp}), \label{65}
\end{equation}
with $g_A>0$.  The self-energy $g_A$ is inversely proportional to the quadrupolar electronic polarizability
and hence $g_{Cs}<g_{Rb}<g_K$ (see Table \ref{Table4}).  These concepts are inspired from the shell model of lattice
dynamics where anisotropic electronic polarizabilities have been introduced \cite{Cow}.  The direct
interchain coupling of alkali quadrupoles is numerically small and will be neglected.  We now consider the
sum
\begin{equation}
V_{ep}^{RR}=V^{Ss}+V^{ss},
\end{equation}
where the subscript $ep$ refers to the quadrupolar electronic polarizability of the cations.  We assume that
the induced cation quadrupoles follow adiabatically the motion
of the C$_{60}^-$ chains.  For a given configuration $\{S_2(\vec{q}_{\perp})\}$ of the latter, we minimize
$V_s^{RR}$ with respect to $s_2(\vec{q}_{\perp})$ and find
\begin{equation}
s_2(\vec{q}_{\perp})=-\frac{1}{2}\frac{\lambda(\vec{q}_{\perp})}{g_A}S_2(\vec{q}_{\perp}).
\end{equation}
Substitution into $V_{ep}^{RR}$ leads to the cation quadrupolar electronic polarizability mediated rotational
interaction
\begin{equation}
V_{ep}^{RR}=-\frac{1}{2}N_1\sum_{\vec{q}_{\perp}}J_{ep}(\vec{q}_{\perp})S_2^{\dagger}(\vec{q}_{\perp})
S_2(\vec{q}_{\perp}), \label{VepRR}
\end{equation}
where
\begin{equation}
J_{ep}(\vec{q}_{\perp})=\frac{1}{2}\frac{\lambda^2(\vec{q}_{\perp})}{g_A}. \label{Jepformule}
\end{equation}
This interaction is always attractive and maximum in absolute value at $\vec{q}_{\perp}=\vec{q}_{\Gamma}$ (as
is the case for the lattice mediated interaction $V_{ac}^{RR}$).  Both $V_{ep}^{RR}$ and $V_{ac}^{RR}$ lead
to a
condensation of $S_2(\vec{q}_{\perp})$ at $\vec{q}_{\Gamma}$ and hence to the ferrorotational structure $I2/m$:
\begin{equation}
\left<S_2(\vec{q}_{\perp})\right>=\sqrt{N_c}\sigma\delta_{\vec{q}_{\perp},\vec{q}_{\Gamma}},
\label{condensation}
\end{equation}
where $\sigma$ is the order parameter amplitude.

In the following section we will discuss the competition between the direct rotational interaction of
C$_{60}$ chains and the indirect, alkali mediated interaction.
\end{subsection}
\end{section}
%
%
%-----------------------
%Stability of Structures
%-----------------------
\begin{section}{Stability of Structures}\label{Section5}
The orientational quadrupolar interaction $V^{RR}$ between C$_{60}$ chains is the sum of the direct
interchain potential $V_{QQ}^{RR}$, Eq.\ (\ref{VRR}) and of the indirect, alkali-mediated potentials $V_{ac}^
{RR}$, Eq.\ (\ref{VacRR}), and $V_{ep}^{RR}$, Eq.\ (\ref{VepRR}).  Hence we write
\begin{equation}
V^{RR}=V_{QQ}^{RR}+V_{ac}^{RR}+V_{ep}^{RR},
\end{equation}
or equivalently
\begin{equation}
V^{RR}=\frac{N_1}{2}\sum_{\vec{q}_{\perp}}J(\vec{q}_{\perp})S_2^{\dagger}(\vec{q}_{\perp})
S_2{\vec{q}_{\perp}},
\end{equation}
with
\begin{equation}
J(\vec{q}_{\perp})=J_{QQ}(\vec{q}_{\perp})-J_{ac}(\vec{q}_{\perp})-J_{ep}(\vec{q}_{\perp}). \label{Jqperp}
\end{equation}
Here the direct interaction $J_{QQ}(\vec{q}_{\perp})$,
due to van der Waals and Coulomb quadrupole-quadrupole potentials,  is given by Eq.\ (\ref{Jq}), with $J_a$ and $J_b$ defined in Eqs.\ (\ref{Jaformule}),
(\ref{Jbformule}).
The acoustic lattice displacements mediated interaction $J_{ac}({\vec{q}_{\perp}})$ is given by
Eq.\ (\ref{Jacformule})
while the electronic polarizability mediated interaction $J_{ep}(\vec{q}_{\perp})$ is defined by
Eq.\ (\ref{Jepformule}).  We have already shown that the direct interaction $J_{QQ}({\vec{q}_{\perp}})$
becomes maximum and attractive
at the Brillouin zone boundary $\vec{q}_{\perp}=\vec{q}_Z$, and hence favors the antiferrorotational
structure $Pmnn$.  On the other hand the alkali mediated interactions $J_{ac}(\vec{q}_{\perp})$ and
$J_{ep}(\vec{q}_{\perp})$ both are attractive and maximum at the Brillouin zone center $\vec{q}_{\perp}=
\vec{q}_{\Gamma}$, and hence favor the ferrorotational structure $I2/m$.  The strength of the indirect
interactions depends
on the specific nature of the alkali ions.

We first consider $J(\vec{q}_{\Gamma})$.  From Eqs.\ (\ref{JQQqgammaformule}), (\ref{Jacqgammaformule}) and
(\ref{Jepformule}) we
find the numerical values quoted in Table \ref{Table5}.

Next we calculate $J(\vec{q}_Z)$.  The results are quoted in Table \ref{Table6}.  
%%%%%%%%%%%%%%%%%%%%%%%%%%%%%%%%%%%%%%%%%%%%%%%%%%%%%%%%%%%%%%%%%%%%%%%%%%%
%Table 6
\begin{table}
\caption{
Interactions at $\vec{q}_{\perp}=\vec{q}_Z$, units K.
\label{Table6}}
\begin{tabular}{cccccc}
& $J_{QQ}^C(\vec{q}_Z)$ & $J_{QQ}^W(\vec{q}_Z)$ & $-J_{ac}(\vec{q}_Z)$ & $-J_{ep}(\vec{q}_Z)$ & $J(\vec{q}_Z)$ \\
\tableline
KC$_{60}$ & $-195.84$ & $-184.70$ & $0$ & $-49.26$ & $-429.80$ \\
%\tableline
RbC$_{60}$ & $-190.21$ & $-176.40$ & $0$ & $-185.80$ & $-552.41$ \\
%\tableline
CsC$_{60}$ & $-188.54$ & $-167.88$ & $0$ & $-227.64$ & $-584.05$ \\
%\tableline
C$_{60}$ & / & $-121.86$ & / & / & $-121.86$ 
\end{tabular}
\end{table}
%%%%%%%%%%%%%%%%%%%%%%%%%%%%%%%%%%%%%%%%%%%%%%%%%%%%%%%%%%%%%%%%%%%%%%%%%%%
We have taken into account that there
is no
coupling to acoustic phonons at $\vec{q}_Z$.  In Fig.\ 3 we have plotted the total rotational interaction
$J(\vec{q}_{\perp})$ as a function of $q_{\zeta}$ along the line $\vec{q}_{\Gamma}-\vec{q}_Z$.  In Eq.\
(\ref{Jqperp}) we have used for $J_{ac}(\vec{q}_{\perp})$ the interpolation expression
\begin{equation}
J_{ac}(\vec{q}_{\perp}=(0,q_{\zeta}))=\frac{8(v_{2,c}^A)^2\sin^2(q_{\zeta}\frac{c}{2})}{q_{\zeta}^2c_{44}abc},
\end{equation}
which coincides with Eq.\ (\ref{Jacqgammaformule}) at $\vec{q}_{\Gamma}$ and vanishes at $\vec{q}_Z$.  While
for KC$_{60}$, with small polarizability of the K$^+$ ion, the direct interchain interaction
$J_{QQ}(\vec{q}_Z)$ dominates and hence leads to $Pmnn$, for RbC$_{60}$ and even more for CsC$_{60}$, the
alkali mediated interaction $J_{ep}(\vec{q}_{\Gamma})$ is most important and leads to the structure $I2/m$.
This structure is monoclinic and indeed one has measured small deviations of the $(\vec{b},\vec{c})$ angle
from $90\;^{\rm o}$ \cite{Rou,Huq}.  The present theory accounts for such shears.  We rewrite expression
(\ref{sdagger0q}) for $\vec{q}_{\perp}=(0,q_{\zeta})$ as
\begin{equation}
q_{\zeta}s_{\eta}^{\dagger}(0,0,q_{\zeta})=\frac{2i\sqrt{m}v_{2,c}^A}{abc_{44}}\sqrt{N_1}S_2^{\dagger}
(0,q_{\zeta}).
\label{qzetasetadagger}
\end{equation}
We observe that
\begin{equation}
\lim_{q_{\zeta}\longrightarrow 0}iq_{\zeta}s_{\eta}(0,0,q_{\zeta})=\sqrt{mN}\epsilon_{\zeta\eta},
\end{equation}
where $\epsilon_{\zeta\eta}$ are the homogeneous shears, and $N=N_1N_c$ is the number of unit cells.  Taking
then the limit $q_{\zeta}\longrightarrow 0$ in Eq.\ (\ref{qzetasetadagger}), we get by using
Eq.\ (\ref{condensation})
\begin{equation}
\epsilon_{\zeta\eta}=\frac{2v_{2,c}^A}{abc_{44}}\sigma
\end{equation}
where $\sigma$ is the order parameter amplitude.  We see that ferrorotational order induces $\zeta$, $\eta$
shears, in accordance with the monoclinic space group $I2/m$.  With the numerical values of $v^A_{2,c}$ from
Table \ref{Table4}, the lattice parameters of Table \ref{Table1} and assuming $\sigma=1$, we obtain for
$\epsilon_{\zeta\eta}$ the values of $0.476\;^{\rm o}$ and $0.483\;^{\rm o}$ for RbC$_{60}$ and CsC$_{60}$
respectively.  The deviations from $90\;^{\rm o}$ of the
$(\vec{b}_o,\vec{c}_o)$ angle measured experimentally are $0.316\;^{\rm o}$ \cite{Huq} and $0.180\;^{\rm o}$
\cite{Rou} for Rb- and CsC$_{60}$ respectively.
\end{section}

%-----------------------------------------------------------------------
%     figure 3
%-----------------------------------------------------------------------
\begin{figure} 
\resizebox{0.46\textwidth}{!}
{ 
 \includegraphics{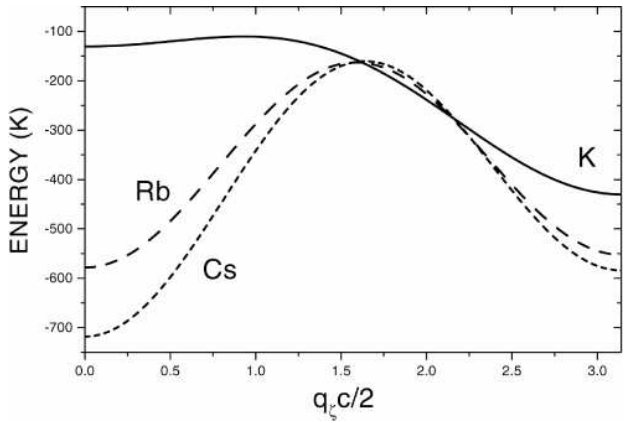} 
} 
\vspace{3mm}
%     figure caption
\caption{
Interchain energy $J(\vec{q}_{\perp})$, units K.
} 
\label{fig3} 
\end{figure} 

%-----------
%Conclusions
%-----------
\begin{section}{Conclusions}\label{Sectionconclusions}
In comparing the numerical values of the various contributions of the total interaction energies in Tables
\ref{Table5} and \ref{Table6} we come to the conclusion that the quadrupolar electronic polarizability
mediated interaction
$J_{ep}(\vec{q}_{\perp})$ and the direct quadrupolar interaction $J_{QQ}(\vec{q}_{\perp})=
J_{QQ}^C(\vec{q}_{\perp})+J_{QQ}^W(\vec{q}_{\perp})$ are the determining factors for the structures $Pmnn$ or
$I2/m$, while the TR mediated interaction $J_{ac}(\vec{q}_{\perp})$ plays quantitatively a negligible role.
The direct quadrupolar interaction, both of van der Waals and Coulomb type, is always attractive and maximum
at $\vec{q}_{\perp}=\vec{q}_Z$ and hence favors the antiferrorotational structure, in agreement with the
space group $Pmnn$ for pressure polymerized C$_{60}$ \cite{Dav2,Mor} and for polymerized KC$_{60}$
\cite{Ste,Lau}.  Notice that the values
of $J_{QQ}(\vec{q}_Z)$ for all three AC$_{60}$ compounds are rather close together, and hence there is no way
to
explain the different structure $I2/m$ for RbC$_{60}$ \cite{Lau,Huq} and CsC$_{60}$ \cite{Rou} by the direct
interchain coupling via C$_{60}$ monomers, the only difference between the three compounds being the lattice
constants.  The role of the alkalis has to be more specific \cite{Mic} than just providing different lattice spacings.

Since the values of $J_{ac}(\vec{q}_{\Gamma})$ are rather small and similar for the three compounds
(Table \ref{Table5}),
the only decisive interaction allowing to discriminate between KC$_{60}$ and RbC$_{60}$, CsC$_{60}$, is the
quadrupolar polarizability of the cations, introduced in Sect.\ \ref{Sectionalkalimediated}.
Here we do not invoke the quadrupolar
polarizability of the inner shells with small ionic radii but the quadrupole formation due to excited
$d$-states \cite{Mic} which carry a residual electronic charge.  The average radii $d_A$ of these states, calculated in
Appendix \ref{AppendixC} and quoted in Table \ref{Table4}, are considerably larger than the ionic radii of K$^+$, Rb$^+$, Cs$^+$ in
conventional compounds like the alkali halides or like the ionic crystals with small molecular ions
(alkali-cyanides, -nitrites).  Within our view the large interstitial space available for the ``octahedral"
alkalis in AC$_{60}$ compounds is a prerequisite for the existence and partial occupancy of these $d$-states
in the solid.  In that respect the alkali-fullerides are unique ionic solids, and we expect that the excited
$d$-states are relevant for an understanding of the electromagnetic properties \cite{For}.  Notice that in the present
approach RbC$_{60}$ and CsC$_{60}$ are similar because their average radii of excited $4d$- and $5d$-states
respectively are close, while
KC$_{60}$ with an excited $3d$-state has a considerably smaller value of $d_A$.  A same trend is also present
in the atomic radii of the transition metals Sc, Y and the inner transition metal La, with $3d^1$, $4d^1$ and
$5d^1$ electrons respectively, with atomic radii $1.64$ {\AA}, $1.82$ {\AA} and $1.88$ {\AA} respectively.

Within the present work we have assumed a model of rigid polymer chains which describes the structures $Pmnn$
and $I2/m$.  The assumption of rigid chains has to be abandoned if one wants to take into account a
modulation of electronic and structural properties along the orthorhombic $\vec{a}$ axis.  In fact the
interaction of electric quadrupoles between C$_{60}$ monomers along a same chain immediately suggests an
antiferro-orientional structure along the chains.  However such an extension is not yet sufficient to
understand the recently discovered superstructure $(\vec{a}+\vec{c},\vec{b},\vec{a}-\vec{c})$ below $T=50$ K
in KC$_{60}$ \cite{Cou}.  Here again the alkalis seem to play a specific role.
\end{section}
\acknowledgements
We acknowledge useful discussions with K.~Knorr, P.~Launois and R.~Moret.  This work has been financially
supported by the Fonds voor Wetenschappelijk Onderzoek, Vlaanderen, and by the Bijzonder Onderzoeksfonds, 
Universiteit Antwerpen, UIA.
%
%
%APPENDICES
%----------
\appendix
%----------
%APPENDIX A
%----------
\begin{section}{}\label{AppendixA}

In the cubic phase the elastic part of the free energy per formula unit AC$_{60}$ reads
\cite{Lan}
\begin{eqnarray}
& & U_{el}=\frac{V_c}{2}\left[c_{11}\left(\epsilon_{xx}^2+\epsilon_{yy}^2+\epsilon_{zz}^2\right)
+2c_{12}\left(\epsilon_{xx}\epsilon_{yy}+\epsilon_{yy}\epsilon_{zz} \right. \right. \nonumber \\
& & +\left. \left. \epsilon_{zz}\epsilon_{xx}\right)
+4c_{44}\left(\epsilon_{xy}^2+\epsilon_{yz}^2+\epsilon_{zx}^2\right)\right].
\end{eqnarray}
Here $V_c=a_c^3/4$ is the volume of the primitive unit cell, $a_c$ is the cubic lattice constant,
$c_{ij}$ are the elastic constants (we use the Voigt notation) and $\epsilon_{ij}$ are the lattice
deformations.  Performing a rotation of the cubic system of axes $(x,y,z)$ to an orthorhombic system
$(\xi,\eta,\zeta)$,
where  $\xi$ corresponds to the cubic $[110]$ direction, $\eta$ to $[\overline{1}10]$ and $\zeta$ to $[001]$,
we get
\begin{eqnarray}
& & U_{el} = \frac{V_c}{2}\left[c_{11}^o\left(\epsilon_{\xi\xi}^2+\epsilon_{\eta\eta}^2\right)
+c_{33}^o\epsilon_{\zeta\zeta}^2
+2c_{12}^o\epsilon_{\xi\xi}\epsilon_{\eta\eta}
+2c_{13}^o 
\right. \nonumber \\
& & \left. \times\ (\epsilon_{\xi\xi}\epsilon_{\zeta\zeta}+\epsilon_{\eta\eta}\epsilon_{\zeta\zeta}) 
+4c_{66}^o\epsilon_{\xi\eta}^2+4c_{55}^o\left(\epsilon_{\xi\zeta}^2+\epsilon_{\eta\zeta}^2\right)\right].
\end{eqnarray}
Here the orthorhombic elastic constants $c_{ij}^o$ are related to the cubic elastic constants by
$c_{11}^o=(c_{11}+c_{12}+2c_{44})/2$, $c_{12}^o=(c_{11}+c_{12}-2c_{44})/2$, $c_{33}^o=c_{11}$,
$c_{23}^o=c_{12}$, $c_{13}^o=c_{12}$, $c_{66}^o=(c_{11}-c_{12})/2$, $c_{55}^o=c_{44}$.  Polymerization
acts as a compression along [110], i. e. along $\xi$.  The corresponding surface forces are described by
the boundary condition \cite{Lan}
\begin{equation}
F_i=\sigma_{ik}n_k,
\end{equation}
where $\sigma_{ik}$ is the elastic stress tensor (indices $i, k$ run over the orthorhombic axes $\xi, \eta,
\zeta$); $\vec{n}$ is the normal to the surface.  Using
\begin{equation}
\sigma_{ik}=\frac{1}{V_c}\frac{\partial U_{el}}{\partial \epsilon_{ik}},
\end{equation}
and observing that $\vec{F}=(-K,0,0)$, where $K>0$ is the strength of the compressional force, we obtain
\begin{mathletters}
\begin{equation}
\sigma_{\xi\xi}=c_{11}^o\epsilon_{\xi\xi}+c_{12}^o\epsilon_{\eta\eta}+c_{13}^o\epsilon_{\zeta\zeta}=-K
\label{sigmaxixi}.
\end{equation}
Since there are no lateral forces, $\sigma_{\eta\eta}=0$ and $\sigma_{\zeta\zeta}=0$, which leads to
\begin{eqnarray}
c_{11}^o\epsilon_{\eta\eta}+c_{12}^o\epsilon_{\xi\xi}+c_{13}^o\epsilon_{\zeta\zeta}=0\label{c11o},\\
c_{33}^o\epsilon_{\zeta\zeta}+c_{13}^o\epsilon_{\xi\xi}+c_{23}^o\epsilon_{\eta\eta}=0\label{c33o},
\end{eqnarray}
\end{mathletters}
and $\sigma_{\xi\eta}=0$, $\sigma_{\zeta\eta}=0$, $\sigma_{\xi\zeta}=0$ lead to
$\epsilon_{\xi\eta}=0$, $\epsilon_{\zeta\eta}=0$, $\epsilon_{\xi\zeta}=0$.  Solving the system of equations
(\ref{sigmaxixi})-(\ref{c33o}) we get the deformations that are quoted in Eqs.\
(\ref{elast1a})-(\ref{elast1c}).
\end{section}
%
%
%----------
%APPENDIX B
%----------
\begin{section}{}\label{AppendixB}
%Born-Mayer and van der Waals Interactions between C60 Monomers
%--------------------------------------------------------------
\begin{subsection}{Born-Mayer and van der Waals Interactions between C$_{60}$ Monomers}
A C$_{60}$ monomer in AC$_{60}$ (A=K, Rb, Cs) is described by a rigid cluster of interaction centers (ICs).
The ICs refer to pair potentials between monomers on separate chains.
In addition to the 60 carbon
atoms, three ICs per double bond (at positions $L/2$ and $\pm L/4$, where $L$ is the bond length) and one IC
per single bond (located at the bond center) are considered.
Following \cite{Lam2} we consider van der Waals potentials of type (\ref{vdWpotential}), where the potential
parameters $C_1$, $C_2$ and $B$ depend on the type of ICs.  Numerical values are given in Table
\ref{TableB1}.  
%%%%%%%%%%%%%%%%%%%%%%%%%%%%%%%%%%%%%%%%%%%%%%%%%%%%%%%%%%%%%%%%%%%%%%%%%%%
%Table B.1
\begin{table}
\caption{
Born-Mayer-van der Waals potential constants for the interactions between the various types of ICs ($a$
stands for ``atom", $db$ for ``double bond IC" and $sb$ for ``single bond IC").  The factors $\frac{1}{9}$
and $\frac{1}{3}$ arise from the threefold multiplicity of the double bond ICs.  For comparison with earlier
work
[34,50],
$C_1$ is expressed in units $3.745\times 10^7$ K and $B$ in units
$3.054\times 10^5$ K$\;${\AA}$^6$, 
$C_2$ is given in {\AA}$^{-1}$. 
\label{TableB1}}

\begin{tabular}{ccccccc}
& $a-a$ & $db-db$ & $sb-sb$ & $a-db$ & $a-sb$ & $db-sb$ \\
\tableline
$C_1$ & $0.864$ & $\frac{1}{9}\times 0.259$ & $0.158$ & $\frac{1}{3}\times 0.169$ & $0.169$ & $0$ \\
%\tableline
$C_2$ & $3.6$ & $3.2$ & $3.6$ & $3.4$ & $3.6$ & $0$ \\
%\tableline
$B$ & $2.1$ & $0$ & $0$ & $0$ & $0$ & $0$
\end{tabular}
\end{table}
%%%%%%%%%%%%%%%%%%%%%%%%%%%%%%%%%%%%%%%%%%%%%%%%%%%%%%%%%%%%%%%%%%%%%%%%%%%
The comparison of experimental X-ray diffuse scattering results with theoretical predictions
based on various intermolecular potentials, shows that such a model has its merits \cite{Lau3}.  However the
transition temperature $T_c$ for the phase change $Fm\overline{3}m\longrightarrow Pa\overline{3}$ obtained by
potentials used in \cite{Lam2} is much too large \cite{Lau3,Mic5} in comparison with the experimental value
$T_c\approx250$ K.  It is then necessary to adjust the potential parameters.  A potential is considered
satisfactory if it reproduces the transition temperature and the experimental crystal field \cite{Cho} in the
orientationally disordered phase of C$_{60}$.  The potential parameters of Table \ref{TableB1} give $T_c=252$
K and crystal field expansion parameters $w_6=346.3$, $w_{10}=-84.1$, $w_{12,1}=-82.9$, $w_{12,2}=-378.6$ for
the disordered phase of solid C$_{60}$.
\end{subsection}
%
%
%Born-Mayer and van der Waals Interactions between a C60 Monomer and an Alkali Atom
%----------------------------------------------------------------------------------
\begin{subsection}{Born-Mayer and van der Waals Interactions between a C$_{60}$ Monomer and an Alkali Atom}
Here we let the alkali atom interact with the 60 carbon atoms of the C$_{60}$ monomer.  We take again the
pair potential of the form of Eq.\ (\ref{vdWpotential}).  Numerical
values for the potential constants are quoted in Table \ref{TableB2}.  
%%%%%%%%%%%%%%%%%%%%%%%%%%%%%%%%%%%%%%%%%%%%%%%%%%%%%%%%%%%%%%%%%%%%%%%%%%%
%Table B.2
\begin{table}
\caption{
Born-Mayer-van der Waals potential constants for the interaction between a carbon atom and an alkali atom.
\label{TableB2}}

\begin{tabular}{ccccc}
& units & C$-$K & C$-$Rb & C$-$Cs \\
\tableline
$C_1$ & $10^7$ K & $2.43$ & $3.59$ & $4.96$ \\
%\tableline
$C_2$ & {\AA}$^{-1}$ & $3.28$ & $3.29$ & $3.32$ \\
%\tableline
$B$ & $10^5$ K$\;${\AA}$^6$ & $3.36$ & $5.26$ & $8.23$
\end{tabular}
\end{table}
%%%%%%%%%%%%%%%%%%%%%%%%%%%%%%%%%%%%%%%%%%%%%%%%%%%%%%%%%%%%%%%%%%%%%%%%%%%
These values are derived from the
homoatomic interaction potential constants: for C$-$C the $a-a$ values of Table \ref{TableB1} were used,
for K$-$K and Rb$-$Rb the values of Ref.\ \cite{Fer} were taken, and for Cs$-$Cs an extrapolation of the
K$-$K and Rb$-$Rb values based on the ratios of the ionic radii of K$^+$, Rb$^+$ and Cs$^+$ was made.
\end{subsection}
\end{section}
%
%
%----------
%APPENDIX C
%----------
\begin{section}{}\label{AppendixC}
To calculate the portion of the charge located inside the interstitial sphere precisely,
one should perform an electron band structure calculation of AC$_{60}$.
However, some conclusions can be made on the basis of a general
consideration.
In the tight-binding approximation we expect that the electron wave function of 
a band electron 
at an alkali site $\vec{n}$, $\psi_{A}(\vec{k},\alpha)|_{\vec{n}}$,
is expanded in terms of local $s$ and $d$ functions, i.e.
\begin{eqnarray} 
 \left. \psi_{A}(\vec{k},\alpha) \right|_{\vec{n}}=\frac{1}{\sqrt{N}}
 e^{i \vec{k} \cdot \vec{X}(\vec{n})} 
 \left[ \gamma_{0\,0}(\vec{k},\alpha) R_s(r)\, Y_0^0 \right.  \nonumber \\
 + \left. \sum_{m=-2}^2 
 \gamma_{2\,m}(\vec{k},\alpha) R_d(r)\, Y_2^m(\hat{r}) \right]  , 
\label{new1}
\end{eqnarray}
where $R_s(r)$ and $R_d(r)$ are the radial parts of $s$ and $d$ states of the 
alkali, respectively. ($N$ is the total number of sites in the crystal.)
Here we are speaking about $4s$ and $3d$ electron states for K, 
$5s$ and $4d$ states for Rb, and $6s$ and $5d$ states of Cs. 
The $s$ and $d$ coefficients $\gamma_{l\, m}(\vec{k},\alpha)$
(i.e. $\gamma_{0\, 0}(\vec{k},\alpha)$ and
$\gamma_{2\,m}(\vec{k},\alpha)$) are found by solving a secular equation
for the band electron with the wave vector $\vec{k}$ and the
band number $\alpha$.
The inclusion of the $d$-states in 
$\psi_{A}(\vec{k},\alpha) |_{\vec{n}}$ is logical
since for a neutral K, Rb and Cs atom the $d$-shell corresponds
to the first excited electron level.
The resulting charge distribution inside the sphere at the site $\vec{n}$ 
can be computed
by summing up the density of all extended states $\vec{k},\alpha$ 
below the Fermi energy, $E(\vec{k},\alpha) \le E_F$.
The charge at the alkali site is small but not zero and we believe that 
$|\gamma_0^0(\vec{k},\alpha)|^2 \sim 0.1$,
$|\gamma_2^m(\vec{k},\alpha)|^2 \sim 0.1$.
It is important that the wave function $\psi_{A}(\vec{k},\alpha)$
has nonzero matrix elements of quadrupolar charge density.
To demonstrate this, we consider quadrupole components of the electronic
density associated with SARFs $S_{\Lambda}(\hat{r})$,
where $\Lambda$ refers either to the two components ($k=1,2$) of $E_g$ symmetry or
to the three components ($k=1,2,3$) of $T_{2g}$ symmetry, i.e.
$\Lambda = (l=2,\Gamma,k)$, where $\Gamma=E_g$ or $T_{2g}$.
In Ref.~\onlinecite{NM} 
it has been shown that the operator of quadrupolar density of
conduction electrons reads
\begin{eqnarray}
 & & \rho_{\Lambda}^{l_1\, l_2}(\vec{q}) =
 {\frac {1}{\sqrt{N}}} \sum_{\alpha, \beta} \sum_{\vec{k}}
 a_{\vec{k} \alpha}^{\dagger} a_{\vec{k}-\vec{q} \beta}\,
 c_{\Lambda \, l_1 l_2} (\vec{k},\alpha; \vec{k}-\vec{q},\beta) ,
   \nonumber \\
 & & \label{new2} 
\end{eqnarray}
where $l_1, \, l_2=0$ or 2 ($s$ and $d$ states), and $a_{\vec{k} \alpha}^{\dagger}$,
$a_{\vec{k} \alpha}$ are creation and annihilation operators for one electron in
the state $(\vec{k},\alpha)$. Here
\begin{eqnarray}
  c_{\Lambda\, l_1 l_2}(\vec{k},\alpha;\,\vec{p},\beta)&=&
  \sum_{m_1, m_2}
  \gamma_{l_1 m_1}^*(\vec{k},\alpha) \, \gamma_{l_2 m_2}(\vec{p},\beta) \nonumber \\ 
  & & \times c_{\Lambda}(l_1 m_1,\,l_2 m_2) .
\label{new3}
\end{eqnarray}
with the quadrupolar matrix elements
\begin{eqnarray}
 c_{\Lambda}(l_1 m_1,\,l_2 m_2)=\int Y_{l_1}^{m_1\,*}(\Omega)\, 
 S_{\Lambda}(\Omega) \, Y_{l_2}^{m_2}(\Omega)\, d\Omega. \nonumber \\
\label{new4}
\end{eqnarray}
Some coefficients $c_{\Lambda}(l_1 m_1,\,l_2 m_2)$ 
are different from zero for certain $d-d$ and $d-s$ transitions
as has been shown in Ref.~\onlinecite{NM} within a more general context
(see there Eq.\ (3.21) for $s-d$ transitions, and 
Table \ref{Table1} for $d-d$ transitions with $\Lambda=(T_{2g},k)$).
From the latter observation we conclude that if there is an
appreciable weight of $d$-states in 
$\psi_{A}(\vec{k},\alpha)|_{\vec{n}}$, Eq.~(\ref{new1}), then the
alkalis acquire a quadrupole moment. (In order to obtain the
operator of the quadrupolar moment for the ground
state of AC$_{60}$ we put $\vec{q}=\vec{k}-\vec{p}=\vec{0}$ and $\alpha=\beta$
in Eqs.\ (\ref{new2}) and (\ref{new3}).)
This consideration gives a microscopic support for the concept of quadrupolar
polarizability of alkalis introduced in Ref.~\onlinecite{Mic}.
It is clear that the quadrupolar moment of alkalis depends on the
coefficients $\gamma_{0\, 0}(\vec{k},\alpha)$,
$\gamma_{2\, m}(\vec{k},\alpha)$. In principle, the secular equation
for the coefficients should also take into account the bilinear
quadrupole-quadrupole interaction with the polymer chains of C$_{60}$
molecules. In practice, however, such calculation would be very
difficult to implement. Instead, following Ref.~\onlinecite{Mic} 
one can introduce
phenomenologically quadrupolar moments of alkalis and then optimize 
their interactions with the neighboring polymer chains.
In this work we use a simple model with two point
charges $q_A$ at distances $\pm d_A$ from the alkali center.
Here $d_A$ is an average $d$-shell radius which is found as
\begin{eqnarray}
 d_A=\langle r \rangle_d = \int_0^{\infty} {\cal R}_d(r')\, {r'}^3 dr' .
\label{new5}
\end{eqnarray}
The radial functions ${\cal R}_d$ and the average radii 
($d_K$, $d_{Rb}$, $d_{Cs}$) were calculated 
numerically. We have employed a relativistic atomic program for
self-consistent-field calculations of K$^+$, Rb$^+$ and Cs$^+$
with the local density approximation (LDA) of exchange according
to Barth-Hedin \cite{BH}. Since the ionicity of alkalis is close to $+1$,
${\cal R}_d (r)$ corresponds to the virtual (almost empty) $3d_{3/2}$ shell 
for K, to the virtual $4d_{3/2}$ level for Rb
and to the virtual $5d_{3/2}$ level for Cs.  The role of excited $d$-functions 
has been discussed by Murrel \cite{Mur}
and by Niebel and Venables \cite{NV} in 
connection with the problem of explaining the observed crystal structure
of the rare gas solids. 

On the other hand, as it has been shown in Ref.\ \onlinecite{NM} there is electronic on-site
interaction which is proportional to the square of the
quadrupolar moment (see Eq.\ (3.14a) of Ref.\ \onlinecite{NM}).
This interaction gives rise to the intraionic self-energy $g_A$
introduced in Eq.\ (6) of Ref.\ \onlinecite{Mic} and Eq.\ (\ref{65}) of the present work.
Upon uniform expansion or
contraction of the electron density around an alkali, 
the strength of it scales as $1/d_A$.
\end{section}


\begin{references}
%1
\bibitem{Dre}
M.S. Dresselhaus, J. Dresselhaus, and P.C. Eklund, {\it Science of Fullerenes and Carbon Nanotubes} (Academic
Press, New York, 1995).

%2
\bibitem{Kuz}
H. Kuzmany, B. Burger, and J. K\"{u}rti, in {\it Optical and Electronic Properties of Fullerenes and
Fullerene-Based Materials}, edited by J. Shinar, Z.V. Vardeny, and Z.H. Kafafi (Marcel Dekker, New York,
2000).

%3
\bibitem{For}
L. Forr\'{o} and L. Mih\'{a}ly, Rep. Progr. Phys. {\bf 64}, 649 (2001).

%4
\bibitem{Win}
J. Winter and H. Kuzmany, Solid State Commun. {\bf 84}, 935 (1992); Q. Zhu, O. Zhou, J.E. Fischer,
A.R. McGhie, W.J. Romanov, R.M. Strongin, M.A. Cichy, and A.B. Smith III, Phys. Rev. B {\bf 47},
13948 (1993).

%5
\bibitem{Pek}
S. Pekker, L. Forr\'{o}, L. Mih\'{a}ly, and A. J\'{a}nossy, Solid State Commun. {\bf 90}, 349 (1994).

%6
\bibitem{Cha}
O. Chauvet, G. Oszl\'{a}nyi, L. Forr\'{o}, P.W. Stephens, M. Tegze, G. Faigel, and A. J\'{a}nossy, Phys. Rev.
Lett. {\bf 72}, 2721 (1994).

%7
\bibitem{Ste}
P.W. Stephens, G. Bortel, G. Faigel, M. Tegze, A. J\'{a}nossy, S. Pekker, G. Oszl\'{a}nyi, and L. Forr\'{o},
Nature {\bf 370}, 636 (1994).

%8
\bibitem{Ren}
B. Renker, H. Schober, F. Gompf, R. Heid, and E. Ressouche, Phys. Rev. B {\bf 53}, 14701 (1996);
B. Renker, H. Schober, and R. Heid, Appl. Phys. A {\bf 64}, 271 (1997).


%9
\bibitem{Rao}
A.M. Rao, P. Zhou, K.-A. Wang, G.T. Hager, J.M. Holden, Y. Wang, W.-T. Lee, X.-X. Bi, P.C. Eklund,
and D.S. Cornett, Science {\bf 259}, 955 (1993).

%10
\bibitem{Bom}
F. Bommeli, L. Degiorgi, P. Wachter, O. Legeza, A. Chauvet, G. J\'{a}nossy, O. Oszl\'{a}nyi, and L. Forr\'{o},
Phys. Rev. B {\bf 51}, 14794 (1995).

%11
\bibitem{All}
H. Alloul, V. Brouet, E. Lafontaine, L. Malier, and L. Forr\'{o}, Phys. Rev. Lett. {\bf 76}, 2922 (1996).

%12
\bibitem{Lau}
P. Launois, R. Moret, J. Hone, and A. Zettl, Phys. Rev. Lett. {\bf 81}, 4420 (1998).

%13
\bibitem{Rou}
S. Rouzi\`{e}re, S. Margadonna, K. Prassides, and A.N. Fitch, Europhys. Lett. {\bf 51}, 314 (2000).

%14
\bibitem{Huq}
A. Huq, P.W. Stephens, G.M. Bendele, and R.M. Ibberson, Chem. Phys. Lett. {\bf 347}, 13 (2001).

%15
\bibitem{Cou}
C. Coulon, A. P\'{e}nicaud, R. Cl\'{e}rac, R. Moret, P. Launois, and J. Hone, Phys. Rev. Lett. {\bf 86},
4346 (2001).

%16
\bibitem{Erw}
S.C. Erwin, G.V. Krishna, and E.J. Mele, Phys. Rev. B {\bf 51}, 7345 (1995).

%17
\bibitem{Tan}
K. Tanaka, T. Saito, Y. Oshima, T. Yamabe, H. Kobayashi, Chem. Phys. Lett. {\bf 272}, 189 (1997).

%18
\bibitem{Aks}
V.L. Aksenov, V.S. Shakmatov, and Y.A. Osipyan, JETP Lett. {\bf 62}, 428 (1995); eidem JETP Lett. {\bf 64},
120 (1996).

%19
\bibitem{Nik}
A.V. Nikolaev, K. Prassides, and K.H. Michel, J. Chem. Phys. {\bf 108}, 4912 (1998).

%20
\bibitem{Mic}
K.H. Michel and A.V. Nikolaev, Phys. Rev. Lett. {\bf 85}, 3197 (2000).

%21
\bibitem{Nik2}
A.V. Nikolaev and K.H. Michel, Solid State Commun. {\bf 117}, 739 (2001).

%22
\bibitem{Lau2}
P. Launois, R. Moret, E. Llusca, J. Hone, and A. Zettl, Synth. Metals {\bf 103}, 2357 (1999).

%23
\bibitem{Dav2}
V.A. Davidov, L.S. Kashevarova, A.V. Rakhmanina, A.V. Dzyabchenko, V.N. Agafonov, P. Dubois, R. Ceolin, and
H. Szwarc, JETP Lett. {\bf 66}, 120 (1997).

%24
\bibitem{Mor}
R. Moret, P. Launois, P.-A. Persson, and B. Sundqvist, Europhys. Lett. {\bf 40}, 55 (1997).

%25
\bibitem{Lan}
L.D. Landau and E.M. Lifshitz, {\it Theory of Elasticity} (Pergamon, New York, 1986).

%26
\bibitem{Sch}
H. Schober, A. T\"{o}lle, B. Renker, R. Heid, and F. Gompf, Phys. Rev. B {\bf 56}, 5937 (1997).

%27
\bibitem{Gue}
H.M. Guerrero, R.L. Cappelletti, D.A. Neumann, and T. Yildirim, Chem. Phys. Lett. {\bf 297}, 265 (1998).

%28
\bibitem{Jam}
H.M. James and T.A. Keenan, J. Chem. Phys. {\bf 31}, 12 (1959).

%29
\bibitem{Pre}
W. Press and A. H\"{u}ller, Acta Crystallograph. A {\bf 29}, 252 (1973).

%30
\bibitem{Yvi}
M. Yvinec and R.M. Pick, J. Phys. (France) {\bf 41}, 1045 (1980).

%31
\bibitem{Pre2}
W. Press, Acta Crystallogr. A {\bf 29}, 257 (1973).

%32
\bibitem{Spr}
M. Sprik, A. Cheng, and M.L. Klein, J. Phys. Chem. {\bf 96}, 2027 (1992).

%33
\bibitem{Hei}
R. Heid, Phys. Rev. B {\bf 47}, 15912 (1993).

%34
\bibitem{Lam2}
D. Lamoen and K.H. Michel, J. Chem. Phys. {\bf 101}, 1435 (1994).

%35
\bibitem{Had}
R.C. Haddon, L.E. Brus, and K. Raghavachari, Chem. Phys. Lett. {\bf 125}, 459 (1986).

%36
\bibitem{deS}
T.M. de Swiet, Y.L. Yarger, T. Wagberg, J. Hone, B.J. Gross, M. Tomaselli, J.J. Titman, A. Zettl,
and M. Mehring, Phys. Rev. Lett. {\bf 84}, 717 (2000).

%37
\bibitem{Lyn}
R.M. Lynden-Bell and K.H. Michel, Revs. Mod. Phys. {\bf 66}, 721 (1994).

%38
\bibitem{YuB}
J. Yu, L. Bi, R.K. Kalia, and P. Vashishta, Phys. Rev. B {\bf 49}, 5008 (1994).

%39
\bibitem{Hul}
A. H\"{u}ller, Z. Physik {\bf 254}, 456 (1972); A. H\"{u}ller and J.W. Kane, J. Chem. Phys. {\bf 61},
3599 (1974).

%40
\bibitem{Kra}
W. Kr\"{a}tschmer, L.D. Lamb, K. Fostiropoulos, and D.R. Huffman, Nature {\bf 347}, 354 (1990).

%41
\bibitem{Cow}
R.A. Cowley, W. Cochran, B.N. Brockhouse, and A.D.B. Woods, Phys. Rev. {\bf 131}, 1030 (1963); R. Migoni, H. Bilz, and D. B\"{a}uerle, Phys. Rev. Lett.
{\bf 37}, 1155 (1976).

%42
\bibitem{Lau3}
P. Launois, S. Ravy, and R. Moret, Phys. Rev. B {\bf 55}, 2651 (1997).

%43
\bibitem{Mic5}
K.H. Michel and J.R.D. Copley, Z. Phys. B {\bf 103}, 369 (1997).

%44
\bibitem{Cho}
P.C. Chow, X. Jiang, G. Reiter, P. Wochner, S.C. Moss, J.D. Axe, J.C. Hanson, R.K. McMullen, R.L. Meng, and
C.W. Chu, Phys. Rev. Lett. {\bf 69}, 2943 (1992); W.I.F. David, R.M. Ibberson, and T. Matsuo, Proc. R. Soc.
London Ser. A {\bf 442}, 129 (1993); P. Schiebel, K. Wulf, W. Prandl, G. Heger, R. Papoular, and W. Paulus,
Acta Crystallogr. A {\bf 52}, 176 (1996).

%45
\bibitem{Fer}
M. Ferrario, I.R. McDonald, and M.L. Klein, J. Chem. Phys. {\bf 84}, 3975 (1986).

%46
\bibitem{NM}
A.V. Nikolaev and K.H. Michel, Eur. Phys. J. B, {\bf 17}, 15 (2000).

%47
\bibitem{BH}
V. von Barth and L. Hedin, J. Phys. C {\bf 5}, 1629 (1972).

%48
\bibitem{Mur}
J.N. Murrell, Discuss Faraday Soc., {\bf 40}, 130 (1965).

%49
\bibitem{NV}
K.F. Niebel and J.A. Venables, in {\it Rare Gas Solids}, M.L. Klein and J.A. Venables eds.,
Academic Press, London, 1976, p. 558.

%50
\bibitem{Nel}
B.J. Nelissen, P.H.M. van Loosdrecht, M.A. Verheijen, A. van der Avoird, and G. Meijer, Chem. Phys. Lett.
{\bf 207}, 343 (1993).

\end{references}
\end{document}